%% file: SnowmassBook-Instrumentation.tex
\documentclass{tcibook}
\usepackage{fancyhea}
\usepackage{work}
\usepackage{bm}       
\usepackage{graphicx}
\usepackage{multirow}
\usepackage{lineno}
\usepackage{threeparttable}
\usepackage[normalem]{ulem}
\usepackage{color}
\usepackage[colorlinks = true,
            linkcolor = blue,
            urlcolor  = blue,
            citecolor = blue,
            anchorcolor = blue]{hyperref}
\usepackage{amssymb}
\usepackage[resetlabels]{multibib}

\input workshopsymbols.tex     

\begin{document}


\pagenumbering{roman}

\parindent=0pt
\parskip=8pt
\setlength{\evensidemargin}{0pt}
\setlength{\oddsidemargin}{0pt}
\setlength{\marginparsep}{0.0in}
\setlength{\marginparwidth}{0.0in}
\marginparpush=0pt


\pagenumbering{arabic}

\renewcommand{\chapname}{chap:intro_}
\renewcommand{\chapterdir}{.}
\renewcommand{\arraystretch}{1.25}
\addtolength{\arraycolsep}{-3pt}

\input Instrumentation.tex

\end{document}

%% file: workshopsymbols.tex
\def\eg{{\it e.g.}}

\def\beq{\begin{equation}}
\def\eeq#1{\label{#1}\end{equation}}
\def\eeqn{\end{equation}}

\newenvironment{Eqnarray}%
   {\arraycolsep 0.14em\begin{eqnarray}}{\end{eqnarray}}
\def\beqa{\begin{Eqnarray}}
\def\eeqa#1{\label{#1}\end{Eqnarray}}
\def\eeqan{\end{Eqnarray}}

\let\bar=\overbar

\def\lsim{\mathrel{\raise.3ex\hbox{$<$\kern-.75em\lower1ex\hbox{$\sim$}}}}
\def\gsim{\mathrel{\raise.3ex\hbox{$>$\kern-.75em\lower1ex\hbox{$\sim$}}}}

\def\del{\partial}
\def\Dslash{\not{\hbox{\kern-4pt $D$}}}
\def\dslash{\not{\hbox{\kern-2pt $\del$}}}
\def\pslash{\not{\hbox{\kern-2pt $p$}}}
\def\ETmiss{\not{\hbox{\kern-4pt $E$}}_T}

\def\Dlr{\mathrel{\raise1.5ex\hbox{$\leftrightarrow$\kern-1em\lower1.5ex\hbox{$D$}}}}

\def\ee{e^+e^-}

\def\MSB{{\bar{M \kern -2pt S}}}
\def\msb{{\bar{\scriptsize M \kern -1pt S}}}

\def\drb{{\bar{\scriptsize D \kern -1pt R}}}



%% file: Instrumentation.tex
\chapter{Report of the Instrumentation Frontier Working Group for Snowmass 2021}

\vspace{1cm}

\begin{center}

{\bf Frontier Conveners:} Phillip~S.~Barbeau$^1$, Petra~Merkel$^2$, Jinlong~Zhang$^3$ \\ \vspace{0.2cm}
{\bf Topical Group Conveners:} Darin~Acosta$^4$, Anthony~A.~Affolder$^5$, Artur~Apresyan$^2$, Gabriella~A.~Carini$^6$, 
Thomas~Cecil$^3$, Amy~Connolly$^7$, C.~Eric~Dahl$^{2,8}$, Allison~Deiana$^9$, Carlos~O.~Escobar$^2$, Juan~Estrada$^2$, 
James~E.~Fast$^{10}$, Maurice~Garcia-Sciveres$^{11}$, Roxanne~Guenette$^{12}$, Kent~Irwin$^{13}$, Albrecht~Karle$^{14}$, 
Wesley~Ketchum$^2$, Reina~H.~Maruyama$^{15}$, F.~Mitchell~Newcomer$^{16}$, John~Parsons$^{17}$, Matthew~Pyle$^{18}$, 
Jennifer~L.~Raaf$^2$, Chris~Rogan$^{19}$, Ian~Shipsey$^{20}$, Bernd~Surrow$^{21}$, Maxim~Titov$^{22}$, Sven~E.~Vahsen$^{23}$, 
Andrew~P.~White$^{24}$, Steven~Worm$^{25,26}$, Minfang~Yeh$^6$, Rachel~Yohay$^{27}$ \\ \vspace{0.2cm}
{\bf Liaisons:} Marina~Artuso$^{28}$, Vallary~Bhopatkar$^{7}$, Steven~D.~Butalla$^{29}$, Katherine~Dunne$^{30}$, Farah~Fahim$^2$, 
Michael~T.~Hedges$^{31}$, Scott~Kravitz$^{11}$, W.~Hugh~Lippincott$^{32}$, Myley~Sanchez$^{27}$, Caterina~Vernieri$^{33}$, 
Jacob~Zettlemoyer$^2$ \\ \vspace{0.2cm}
$^1$~Duke University, \\ 
$^2$~Fermi National Accelerator Laboratory, \\
$^3$~Argonne National Laboratory, \\
$^4$~Rice University, \\
$^5$~University of California Santa Cruz, \\
$^6$~Brookhaven National Laboratory, \\
$^7$~Ohio State University, \\
$^8$~Northwestern University, \\
$^9$~Southern Methodist University, \\
$^{10}$~Jefferson Lab, \\
$^{11}$~Lawrence Berkeley National Laboratory, \\
$^{12}$~University of Manchester, \\
$^{13}$~Stanford University, \\
$^{14}$~University of Wisconsin Madison, \\
$^{15}$~Yale University, \\
$^{16}$~University of Pennsylvania, \\
$^{17}$~Columbia University, \\
$^{18}$~University of California Berkeley, \\
$^{19}$~University of Kansas, \\
$^{20}$~Oxford University, \\
$^{21}$~Temple University, \\
$^{22}$~IRFU, CEA, Universit\'e Paris-Saclay, \\
$^{23}$~University of Hawaii Honolulu \\
$^{24}$~University of Texas Arlington, \\
$^{25}$~Deutsches Elektronen-Synchrotron, \\
$^{26}$~Humboldt-Universit\"at zu Berlin, \\
$^{27}$~Florida State University, \\
$^{28}$~Syracuse University, \\
$^{29}$~Florida Institute of Technology, \\
$^{30}$~Stockholm University, \\
$^{31}$~Purdue University, \\
$^{32}$~University of California Santa Barbara, \\
$^{33}$~Stanford Linear Accelerator Laboratory

\end{center}

\setlength{\parindent}{15pt}\indent Detector instrumentation is at the heart of scientific discoveries. Cutting edge technologies enable US particle physics to play a 
leading role worldwide. This report summarizes the current status of instrumentation for High Energy Physics (HEP), the challenges and 
needs of future experiments and indicates high priority research areas. The Instrumentation Frontier studies detector technologies 
and Research and Development (R\&D) needed for future experiments in collider physics, neutrino physics, rare and precision physics 
and at the cosmic frontier. It is divided into more or less diagonal areas with some overlap among a few of them. We lay out five 
high-level key messages that are geared towards ensuring the health and competitiveness of the US detector instrumentation 
community, and thus the entire particle physics landscape. 
\setlength{\parindent}{0pt}

\section{Executive Summary}
While the fundamental science questions addressed by HEP have never been more compelling, there is acute awareness of the challenging 
technical constraints when scaling current technologies. Furthermore, many technologies are reaching their sensitivity limit and new 
approaches need to be developed to overcome the currently irreducible technological challenges and to cope with ever increasing 
extreme data volumes. For the field of HEP to continue to have a bright future, priority within the field must be given to investments 
in both evolutionary and transformational detector development that is coordinated across the National Labs and with the 
university community, international partners, other disciplines and industry. This 
situation is unfolding against a backdrop of declining funding for instrumentation, both at the National Labs and in 
particular at the universities. This trend has to be reversed for the country to continue to play a leadership role in particle 
physics. In this challenging environment it is essential that the community invest anew in instrumentation and optimize the use of 
the available resources to develop new innovative, cost-effective instrumentation. We need to invest in modernized facilities with 
enhanced capabilities, address intermediate planned project needs, and carry out blue-sky R\&D that enable new physics opportunities 
to successfully accomplish the mission of HEP. Many of the findings of this Snowmass report are still well aligned with the 
2019 Report of the Office of Science Workshop on Basic Research Needs for HEP Detector Research and Development (BRN 
report)~\cite{osti_1659761}, but some additions and updates have been identified.

Much of the HEP portfolio is dominated by mid-sized to large scale experiments, which take decades to design, 
build and operate, and often span the entire lifetime of individuals' careers. In this environment it is becoming increasingly 
difficult for students and young researchers to have the opportunity to learn hands-on about detector instrumentation. For example 
on the collider physics side, where in addition to shortages in coverage of detector operations expertise, there will be potentially 
a large gap in time between the current High Luminosity Large Hadron Collider (HL-LHC) upgrades finishing and concrete designs and 
construction for the next generation collider experiments starting. The key challenge here will be to maintain the technical skills 
and experience within the community, and to train a new generation of instrumentation experts. This gap will have to be bridged 
partially by an increased effort in generic detector R\&D, as well by encouraging and supporting Primary Investigators (PI) to get 
involved in other short-term, smaller scale experiments in other areas, such as astro-particle physics or accelerator-based dark 
matter and rare processes experiments. The HEP community should encourage sufficient funding for small-scale experiments. An 
expansion into newly enabled Quantum Information Science (QIS) experiments, exploiting recent transformative technological advances 
should also be encouraged. In general, careers in detector instrumentation in HEP need better support. A valuable addition would be 
increased opportunities for interdisciplinary PhD, leveraging the Science Graduate Student Research (SCGSR) program. Furthermore, 
the field needs more mid-career positions for instrumentation experts, especially at Universities, where an additional advantage 
lies in the potential for cross-cutting positions with other departments, such as material science, quantum information, or 
engineering disciplines. When reviewing such mid-career and even permanent positions for primarily instrumentation physicists, a 
different set of metrics might have to be taken into account in order to evaluate skills, achievements and future potential. 

In addition to the maintenance of the technical workforce and expertise, it is also crucial that the community invests in modern 
facilities, which enable breakthrough advances in detector technologies for future experiments. A coherent set of supporting facilities 
needs to be maintained and enhanced over the coming years. Furthermore, the US HEP community needs to consider the creation of a detector 
R\&D collaborative framework similar or connected to the successor of the RD collaboration model at the European Center for Nuclear 
Research (CERN), which is consolidating detector R\&D collaborations under the European Committee for Future Accelerators (ECFA). These 
could be in specific technological areas and under the guidance and oversight of the Coordinating Panel for Advanced Detectors (CPAD) of 
the American Physical Society's Division of Particles and Fields (APS/DPF). 

In order to stimulate transformational breakthroughs, we need to pursue synergies with other disciplines outside of HEP, as well as 
close collaborations with industry. Recent successful examples of this are QIS and Microelectronics, where communication and exchange 
between the HEP and other communities led to enhanced funding for technology development, which together with a new suite of available 
or achievable detector technologies and methods opened up a new suite of small-scale experiments that enhance the HEP portfolio, and 
can in turn contribute new solutions to traditional experiments. 

To conclude, the Snowmass science frontiers, Cosmic Frontier, Energy Frontier, Neutrino Frontier and Rare and Precision Frontier, 
are pursuing a breadth of different future experiments, which have aggressive schedules and are technically challenging. They 
urgently require significant developments within the instrumentation frontier, which in turn requires significant investment now. 
The key challenges can be summarized as follows:
\begin{itemize}
 \item[\bf IF-1] {Advance performance limits of existing technologies and develop new techniques and materials, nurture enabling 
 technologies for new physics, and scale new sensors and readout electronics to large, integrated systems using co-design methods.}
 \item[\bf IF-2] {Develop and maintain the critical and diverse technical workforce, and enable careers for technicians, engineers 
 and scientists across disciplines working in HEP instrumentation, at laboratories and universities.}
 \item[\bf IF-3] {Double the US Detector R\&D budget over the next five years, and modify existing funding models to enable R\&D 
 consortia along critical key technologies for the planned long term science projects, sustaining the support for such collaborations 
 for the needed duration and scale.}
 \item[\bf IF-4] {Expand and sustain support for blue-sky R\&D, small-scale R\&D, and seed funding. Establish a separate agency 
 review process for such pathfinder R\&D, independently from other research reviews.}
 \item[\bf IF-5] {Develop and maintain critical facilities, centers and capabilities for the sharing of common knowledge and tools, 
 as well as develop and maintain close connections with international technology roadmaps, other disciplines and industry.}
\end{itemize}

In the following we summarize detailed findings and needs from the ten Topical Groups of the Instrumentation Frontier. Further details 
are given in the individual Topical Group reports.

\section{Quantum Sensors}
The use of quantum sensors in HEP has seen explosive growth since the previous Snowmass Community Study. This growth 
extends far beyond HEP impacting many areas of science from communications to cryptography to computing. Quantum sensors have been 
used in searches for dark matter - particle and wave, fifth forces, dark photons, permanent electric dipole moment (EDM), variations 
in fundamental constants, and gravitational waves, among others. These sensors come in a wide range of technologies: atom 
interferometers and atomic clocks, magnetometers, quantum calorimeters and superconducting sensors to name a few. Early work with 
quantum sensors in the context of particle physics often focused in cosmic and rare and precision frontiers, but recent concepts seek 
to expand the use of quantum sensors to the energy and neutrino frontiers solidifying them as fundamental technologies for the future 
of experimental HEP. Based upon input to the Snowmass process and outlined in our Topical Group report~\cite{Cecil2022}, our Topical 
Group has identified several key messages necessary to support the development and use of quantum sensors in HEP:

 \begin{itemize}
 \setlength{\itemindent}{1.5em}
 \item[\bf IF01-1] \textbf{Continue strong support for a broad range of quantum sensors. Quantum sensors address scientific needs 
 across several frontiers and different technologies carve out unique parameter spaces.} While these sensors share many common 
 characteristics, each has advantages that make it the sensor of choice for specific applications along with challenges that need 
 further development to make the greatest impact. 
 \item[\bf IF01-2] \textbf{Continue support for R\&D and operation of table-top scale experiments. Many are shovel ready and have 
 the potential for large impact.} Much of the growth in quantum sensors over the past decade has occurred in small, laboratory 
 based experiments. These fast-paced small experiments should continue to be supported as a way to rapidly develop sensor technology 
 and help determine those areas where quantum sensors can have the greatest impact.
 \item[\bf IF01-3] \textbf{Balance support of tabletop experiments with pathfinders R\&D to address the large-scale challenges of 
 scaling up experiments which will require National Lab and HEP core competences.} As the fast-paced, small experiments mature, those 
 with significant discovery potential begin to emerge along with areas of commonality between the experiments (e.g.\ the need for 
 advanced high field magnets for axion dark matter experiments or ultra-stable lasers for atom interferometers and clocks). They have 
 reached the point at which plans for larger-scale, longer-term experiments should be conceptualized. These concepts can evaluate the 
 potential reach that can be achieved in a larger effort and the scale of require technological development. 
 \item[\bf IF01-4] \textbf{Develop mechanisms to support interactions outside of the HEP program to enable collaborations with fields 
 with developed expertise in quantum sensors. Advances in QIS provide exceptional theoretical and experimental resources to advance 
 quantum sensing that could provide mutual benefits in several areas such as materials, detectors, and devices.} Many of the most 
 promising quantum sensors for HEP science have been developing for the past decade or more in areas outside of the traditional HEP 
 science and funding sphere. For example, atomic clocks developed over many decades as a source of precision timing standards are now 
 stable enough they can be used in the search for variations of fundamental constants and gravitational waves. The HEP community should 
 strive to collaborate with these broader communities in a way the gives HEP access to new sensor technologies while sharing HEP 
 expertise (e.g.\ large magnets and vacuum systems). Effort should be made to allow the free flow of ideas and effort across 
 traditional funding boundaries to encourage scientists and engineers working with quantum sensors to tackle the most interesting 
 and challenging problems available. 
 \item[\bf IF01-5] \textbf{Develop mechanisms to facilitate interactions to support theoretical work address issues of materials and 
 measurement methods.} As with other instrumentation frontiers and as quantum sensors become more sensitive, focused support on 
 quantum materials at the interface of quantum sensors and HEP will be needed. This includes theoretical work necessary for on topics 
 including quantum materials, squeezing, and back action.
 \item[\bf IF01-6] \textbf{Workforce development is needed to encourage workers with the needed skills to engage with the HEP field, 
 maintain current momentum, and for long-term success in the face of growing competition from industrial quantum computing.} While the 
 HEP community is poised to benefit from quantum sensor developments outside of HEP, we face a shortage of skilled 
 workers. The explosive growth in quantum computing in recent years, along with arrival of several major tech companies has created a 
 fierce competition for workers with skills needed to develop quantum sensors and experiments. The HEP community will need to invest 
 now in order to train and retain the next generation of quantum scientist. Increasing collaborations outside of HEP -- as discussed 
 above -- can provide an additional pathway to reaching skilled workers and engaging them on HEP challenges.
\end{itemize}

\section{Photon Detectors}
The Photon Detectors Topical Group has identified two areas where focused R\&D over the next decade could have a large impact in HEP 
experiments. These areas described here are characterized by the convergence of a compelling scientific need and recent 
technological advances. A short summary can be found below, while a more extensive report, including a large number of references for 
further reading, is given in the Photon Detector Topical Group report~\cite{IF02-report}.

The development of detectors with the capability of counting single photons from infrared (IR) to ultra-violet (UV) has been a very 
active area in the last decade. Demonstration for sensors based on novel semiconductor technologies such as Complementary 
Metal-Oxide-Semiconductor (CMOS), Charge Coupled Devices (CCD), skipper-CCDs, and Silicon Photomultiplier (SiPM) as well as 
superconducting technologies such as Microwave Kinetic Inductance Detectors (MKID), Superconducting Nanowire Single Photon Detectors 
(SNSPD) and Transition Edge Sensors (TES) have been performed. These sensors open a new window for HEP experiments in the low photon 
number regime. Several ongoing and future projects in HEP benefit from these developments (Cosmology, Dark Matter, Neutrinos) which 
will also have a large impact outside HEP such as in Basic Energy Science (BES), QIS, and Astronomy. The combined scientific needs 
and technological opportunities make photon counting an ideal area for focused R\&D investment in the coming decade. R\&D is needed 
for enabling large arrays of the new sensors, improving their energy resolution, timing, dark counts rates and extending their 
wavelength coverage. Such investment will secure leadership in photon counting technologies in the US, producing a large impact in HEP, 
with applications outside HEP. This opportunity is summarized in Ref~\cite{PhotonCountingWP}.

A technological solution for the photon detection system in the first two modules of the DUNE 
detector exist, based on the by now well-known Arapuca light traps, with wavelength shifters and SiPMs as the photon detector. However, 
new photon detector developments are being considered for modules 3 and 4. Some of the proposed ideas consist on novel light collectors, 
the so-called dichroicons which are Winston-style light concentrators made from dichroic mirrors, allowing photons to be sorted by 
wavelength, directing the long-wavelength end of broad-band Cherenkov light to photon sensors that have good sensitivity to those 
wavelengths, while directing narrow-band short-wavelength scintillation light to other sensors. This technology could be used in 
water-based liquid scintillators thus realizing a hybrid Cherenkov and scintillator detector. 

One significant development in the last decade is the commercialization of Large Area Picosecond Photodetectors (LAPPDs) and their first 
deployment in a neutrino experiment, ANNIE. LAPPDs are imaging single-photon 
sensors, capable of resolving photon hit positions with sub-cm resolution and arrival times of around 50 ps. LAPPDs bring capabilities 
that may make them useful in future liquid Argon (LAr) detectors, and as part of the DUNE near detector system, particularly the gaseous 
Argon near detector (ND-GAr). They also have the requisite time resolution to separate between early-arriving Cherenkov light and slower 
scintillation light.

Also notable are research and development efforts in new materials that could be directly sensitive to the vacuum ultra-violet (VUV) 
light, such as amorphous selenium (a-Se) and organic semiconductors. R\&D is needed to move from concept demonstrations to full scale 
implementations in an HEP experiment. Future investment in the above technologies over the next decade will enhance the science of the 
DUNE project.

The science for generation, detection and manipulation of light is extremely fast moving and driven mainly from outside our field. The 
HEP community would benefit from resources dedicated to the implementation of these advanced photonic technologies in particle physics 
experiments. 

\begin{itemize}
\setlength{\itemindent}{1.5em}
 \item[\bf IF02-1] The development of detectors with the capability of counting single photons from IR to UV has been a very active 
area in the last decade. We now need to pursue R\&D to implement these in HEP experiments by making larger arrays, improving their 
energy resolution, timing, dark counts rates and extending their wavelength coverage. 

\item[\bf IF02-2] New photon detector developments are being considered for planned future neutrino experiments going beyond the 
current technologies. Concept demonstrations have been done, and we now need to move from a conceptual phase to working detectors.
\end{itemize}

\subsection{Photon Counting Sensors Enabling HEP}
Several novel cosmological facilities for wide-field multi-object spectroscopy were proposed for the Astro2020 decadal review, and are 
being considered as part of the Snowmass process. Ground-based spectroscopic observations of faint astronomical sources in the 
low-signal, low-background regime are currently limited by detector readout noise. Significant gains in survey efficiency can be 
achieved through reductions by using sensors with readout noise below 1e-. Pushing the current photon counters in the direction of 
mega pixel arrays with fast frame rate (about 10 fps) would make this possible.

Dark matter searches based on photon counting technologies currently hold the world record sensitivity for low mass electron-recoil 
dark matter semiconductors are among the most promising detector technologies for the construction of a large multi-kg experiment for 
probing electron recoils from sub-GeV dark matter (DM) (skipper-CCDs). Significant R\&D is needed to scale the experiments from the 
relatively small pathfinders to the multi-kg experiments in the future. 

Single photon counting sensors have also gained importance for their potential as Coherent Elastic neutrino Nucleus Scattering (CEvNS)
detectors. The deposited energy from CEvNS is less than a few keV. Only part of this energy is converted into detectable signal in 
the sensor (ionization, phonons, etc.) and therefore low threshold technologies are needed. Semiconductor and superconducting 
technologies with eV and sub-eV energy resolution for photon counting capability in the visible and near-IR are natural candidates to 
reach the necessary resolution for this application.

Photon counting also enables a wide range of science outside HEP including QIS, BES and applications in radiation detection.

Recent advances in photon counting technology are discussed here. The advances can be grouped in three areas. 

\subsubsection{Superconducting Sensors}
\paragraph{MKIDs} work on the principle that incident photons change the surface impedance of a superconductor through the kinetic 
inductance effect. The magnitude of the change in surface impedance is proportional to the amount of energy deposited in the 
superconductor, allowing for single photon spectroscopy on chip. Frequency multiplexed arrays of 20,440 pixels with energy resolution 
R=E/$\Delta$E$\sim$9.5 at 980 nm, and a quantum efficiency (QE) of ${\sim}$35\% have been achieved. R\&D focused on 
larger arrays with higher QE and better energy resolution would address the needs of the community.

\paragraph{TES} is a photon detector, which utilizes a patterned superconducting film with a sharp superconducting-to-resistive 
transition profile as a thermometer. It is a thermal detector with a well developed theoretical understanding. When a visible or 
infrared photon is absorbed by a TES, the tiny electromagnetic (EM) energy of the photon increases the temperature of the TES and therefore 
changes its resistance. TES have been developed to measure single photons in quantum communication, for axion-like particle searches, 
direct detection of dark matter particles and astrophysical observations in the wavelengths between UV and IR. TES 
detectors can be multiplexed enabling arrays of large channel counts. Multiplexers for detector arrays using 16,000 TES have already 
been successfully implemented. R\&D exploiting microwave resonance techniques has the potential to increase the multiplexing capacity 
by another factor of ten. 

\paragraph{SNSPDs} consist of a superconducting film patterned into a wire with nanometer scale dimensions (although recently devices 
with micrometer-scale widths have been shown to be single-photon sensitive). SNSPDs have been reported with single photon sensitivity 
for wavelengths out to several microns, timing jitter as low as a few~ps, dark count rates down to $6\times10^{-6}$ Hz, and detection 
efficiency of 0.98. They have also been shown to function in magnetic fields of up to 6T. Current R\&D consist on scaling to large 
arrays and extending the spectral range for these sensors to address the needs of HEP and Astrophysics.

\subsubsection{Semiconducting Sensors}
\paragraph{Skipper-CCDs} have an output readout stage that allows multiple non-destructive sampling of the charge packet in each pixel 
of the array thanks to its floating gate output sense node. This non-destructive readout has been used to achieve deep sub-electron 
noise in mega-pixel arrays. Skipper-CCDs fabricated on high resistivity silicon have also demonstrated an extremely low production of 
dark counts. This technology has motivated a new generation of DM and neutrino experiments. The R\&D effort here is currently focused 
on faster readout (10 fps) and large gigapixel arrays.

\paragraph{CMOS} The down scaling of CMOS technology has allowed the implementation of pixels with a very low capacitance, and 
therefore, high sensitivity and low noise (1-2\,$\rm e^-$) at room temperature and high frame rates (50-100\,fps). Commercial cameras 
with sub-electron noise at 5\,fps are now available. These sensors have not yet played a big role in HEP mainly because of the small 
pixel size. Active R\&D taking advantage of the CMOS fabrication process to address the needs of HEP is ongoing, including the 
development of new CMOS sensors with non-destructive readout (skipper-CMOS). These sensors could address the readout speed limitations 
of other semiconductor photon counters. The single photon avalanche diode (SPADs) have also been implemented in standard CMOS 
technology and integrated with on-chip quenching and recharge circuitry addressing fast timing and radiation tolerance requirements 
from HEP.

\paragraph{Photon-to-Digital Converters (PDC)} In a PDC, each SPAD is coupled to its own electronic quenching circuit. This one-to-one 
coupling provides control on individual SPADs and signals each detected avalanche as a digital signal to a signal processing unit 
within the PDC. Hence, PDCs provide a direct photon to digital conversion considering that intrinsically a SPAD is a Boolean detector 
by design. Digital SiPMs were first reported in 1998 by the MIT Lincoln Lab (MIT-LL) and many contributions followed. A major step came 
with microelectronics integration to fabricate both the SPAD and readout quenching circuit in a single commercial process. These 
innovations led to the first multi-pixel digitally read SPAD arrays. A recent review can be found in Ref.~\cite{Pratte2021_Sensors}. 

\subsubsection{Extending Wave Length Coverage}
As has been pointed out also in~\cite{https://doi.org/10.48550/arxiv.2203.08297}, Sec.3.3.3, low-gap materials are being developed 
which could also serve as THz-IR single photon detectors.

\paragraph{Germanium Semiconductors}
Silicon CCDs are commonly utilized for scientific imaging applications in the visible and near infrared. These devices offer numerous 
advantages described previously, while the skipper-CCD adds to these capabilities by enabling multiple samples during readout to reduce 
read noise to negligible levels. CCDs built on bulk germanium offer all of the advantages of Silicon CCDs while covering an even broader 
spectral range. The R\&D in this area will extend the photon counting capabilities of semiconductor into the IR.

\paragraph{UV}
It has been shown that CCD sensitivity can be increased close to the reflection-limited quantum efficiency of silicon. This was done by 
blocking the surface fields and traps through the epitaxial growth of a strongly doped very thin silicon layer (delta-doping). QE 
exceeding 50\% was demonstrated in CCDs down to 125~nm wavelength. The method was demonstrated efficiently on backside-illuminated 
SPAD-based detectors. Other methods to address the surface field and trap issues were also demonstrated. Work is being done to 
enhance SiPMs for the detection of the fast scintillation component of Barium Fluoride (BaF$_2$). An extensive study of the 
delta-doping approach to enhance VUV sensitivity in frontside-illuminated SPAD-based detectors has also been carried out.

\subsection{Photon Detectors For Neutrino Experiments}
A large number of outstanding questions remain to the fundamental nature of the neutrino, which can be probed through the use of higher 
energy ($\mathcal{O}$(MeV) $< E < \mathcal{O}$(GeV)) neutrino sources (\eg, accelerator and atmospheric neutrinos). The nature of these 
remaining puzzles break into the distance over which the neutrinos are allowed to propagate before being detected. Thus the future class 
of experiments are classified as ``short-baseline'' and ``long-baseline'' experiments. 

The next generation long-baseline neutrino experiments aim to answer the questions of the exact ordering of the neutrino mass states, 
known as the mass hierarchy, as well as the size of the CP-violating phase $\delta$. These, as yet unknown quantities, remain one of the 
last major pieces of the Standard Model of particle physics and offer the opportunity to answer such fundamental questions as ``what is 
the origin of the matter/antimatter asymmetry in the universe?'' and ``do we understand the fundamental symmetries of the universe?''. 
By measuring the asymmetry between appearance of electron neutrinos from a beam of muon neutrinos ($P(\nu_{\mu} \rightarrow \nu_{e}$)) 
compared to the appearance of electron antineutrinos from a beam of muon antineutrinos and $P(\bar{\nu}_{\mu} \rightarrow \bar{\nu}_{e}$)) 
as well as the precise measurement of the $\nu_{e}$ energy spectrum measured at the far detector, both the CP violating phase 
($\delta_{CP}$) and the mass hierarchy can be measured in the same experiment.

The Short-Baseline Neutrino (SBN) program aims to address the anomalous neutrino results seen by the Liquid Scintillator Neutrino 
Detector (LSND) and MiniBooNE experiments, which suggest the possible existence of an eV mass-scale sterile neutrino. However, the 
experimental landscape is perplexing since a number of other experiments utilizing a range of different neutrino sources which 
should have been sensitive to such a sterile neutrino have observed only the standard three neutrino oscillations. In this 
landscape, the conclusive assessment of the experimental hints of sterile neutrinos becomes a very high priority for the field of 
neutrino physics.

To address both of these areas of neutrino research, large scale noble element time projection chambers (TPC) play a central role and 
offer an opportunity to perform discovery level measurements through the enhancement of their capabilities. In a noble element TPC, 
particles interact with the medium and deposit their energy into three main channels: heat, ionization, and scintillation light. 
Depending on the physics of interest, noble element detectors attempt to exploit one or more of these signal components. Liquid Noble 
TPCs produce ionization electrons and scintillation photons as charged particles traverse the bulk material. An external electric 
field allows the ionization electrons to drift towards the anode of the detector and be collected on charge sensitive readout or 
transform energy carried by the charge into a secondary pulse of scintillation light. 

A technological solution for the photon detection system in the first two modules of the DUNE detector exist, based on the Arapuca 
light traps, with wavelength shifters and SiPMs as the photon detector. This approach is also being used in other SBN oscillation 
experiments, but not exclusively. Beyond this, new photon detector developments are being considered for DUNE modules 3 and 4 as 
well as for other approved or proposed experiments. 

\subsubsection{Going Beyond DUNE's First Two Modules}
Quoting from the executive summary of the Snowmass IF02 White Paper\emph{Future Advances in Photon-Based Neutrino 
Detectors}~\cite{Klein}, large-scale, monolithic detectors that use either Cherenkov or scintillation light have played major roles 
in nearly every discovery of neutrino oscillation phenomena or observation of astrophysical neutrinos. New detectors at even larger 
scales are being built right now, including Jiangmen Underground Neutrino Observatory (JUNO), Hyper-Kamiokande (Hyper-K), and DUNE. 
These new technologies will lead to neutrino physics and astrophysics programs of great breadth: from high-precision accelerator 
neutrino oscillation measurements, to detection of reactor and solar neutrinos, and even to neutrinoless double beta decay 
measurements that will probe the normal hierarchy regime. They will also be valuable for neutrino applications, such as 
non-proliferation via reactor monitoring.

Of particular community interest is the development of hybrid Cherenkov-scintillation detectors, which can simultaneously exploit 
the advantages of Cherenkov light's reconstruction of direction and related high-energy particle identification (PID) and the 
advantages of scintillation light, high light-yield, low-threshold detection with low-energy PID. Hybrid Cherenkov-scintillation 
detectors could have an exceptionally broad dynamic range in a single experiment, allowing them to have both high-energy, 
accelerator-based sensitivity while also achieving a broad low-energy neutrino physics and astrophysics program. Recently the 
Borexino Collaboration has published results showing that even in a detector with standard scintillator and no special photon sensing 
or collecting, Cherenkov and scintillation light can be discriminated well enough on a statistical basis that a sub-MeV solar 
neutrino direction peak can be seen. Thus the era of hybrid detectors has begun, and many of the enabling technologies described 
here will make full event-by-event direction reconstruction in such detectors possible.

\paragraph{LAPPDs}
New advances in the science of photomultiplier tubes, including long-wavelength sensitivity, and significant improvements in timing 
even with devices as large as 8 inches, make hybrid Cherenkov-scintillation detectors even better, with high light yields for both 
Cherenkov and scintillation light with good separation between the two types of light. LAPPDs have pushed photon timing into the 
picosecond regime, allowing Cherenkov-scintillation separation to be done even with standard scintillation time profiles. The fast 
timing of LAPPDs also makes reconstruction of event detailed enough to track particles with the produced photons. The precision 
timing capabilities of LAPPDs make possible new stroboscopic techniques where different energy components of a wide-band neutrino 
beam can be selected based on the arrival of the neutrinos relative to the beam Radio Frequency (RF) timing. Future neutrino 
experiments, including possible upgrades to the Long Baseline Neutrino Facility (LBNF), could enable sufficiently short proton 
bunches to make this technique viable. At the detector end, this technique requires time resolutions on the order of no more than 
a few hundred picoseconds. Hybrid Cherenkov-scintillation detectors with LAPPDs are naturally well suited for this level of vertex 
precision. The continued use and development of application readiness of LAPPDs in experiments like ANNIE is critical to their 
future viability. Additional investments in R\&D to further grow production yields and reduce costs will also benefit future 
large-scale HEP efforts. 

\paragraph{Dichroicons}
Dichroicons, which are Winston-style light concentrators made from dichroic mirrors. Winston cones are non-imaging light concentrators 
intended to funnel all wavelengths passing through the entrance aperture out through the exit aperture. In turn, a dichroicon allows 
photons to be sorted by wavelength thus directing the long-wavelength end of broad-band Cherenkov light to photon sensors that have 
good sensitivity to those wavelengths, while directing narrow-band short-wavelength scintillation light to other sensors. Dichroicons 
are particularly useful in high-coverage hybrid Cherenkov-scintillation detectors. 

\section{Solid State Detectors and Tracking}
Tracking detectors are of vital importance for collider-based HEP experiments. The primary purpose of tracking detectors is the 
precise reconstruction of charged particle trajectories and the reconstruction of secondary vertices. The performance requirements 
from the community posed by the future collider experiments require an evolution of tracking systems, necessitating the development 
of new techniques, materials and technologies in order to fully exploit their physics potential. 

Technological developments currently underway aim to address these issues, and the successful completion of the programs outlined below 
requires focused efforts from the community on the steady development and refinement of existing technologies, and the pursuit of novel 
blue-sky technologies to enable transformative progress. The HEP community gathered at Seattle Snowmass Summer Meeting in 2022 
identified the following key directions for the near-term priorities of the solid-state tracking. The full report of this Topical Group 
can be found in~\cite{https://doi.org/10.48550/arxiv.2209.03607}.

\begin{itemize}
 \setlength{\itemindent}{1.5em}
 \item [\bf IF03-1] Develop high spatial resolution pixel detectors with precise per-pixel time resolution to resolve individual 
 interactions in high-collision-density environments
 \item [\bf IF03-2] Adapt new materials and fabrication/integration techniques for particle tracking in harsh environments, including 
 sensors, support structures and cooling
 \item [\bf IF03-3] Realize scalable, irreducible-mass trackers in extreme conditions
 \item [\bf IF03-4] Push advanced modeling for simulation tools, developing required extensions for new devices, to drive device 
 design.
 \item [\bf IF03-5] Provide training and retain expert workforce to enable future tracking systems to be designed, developed, 
 constructed and simulated.
 \item [\bf IF03-6] Nurture collaborative networks, provide technology benchmarks and roadmaps and funding in order to develop 
 required technologies on necessary time scales, costs and scope.
\end{itemize}

Advanced 4-dimensional (4D) trackers with ultra-fast timing (10-30~ps) and extreme spatial resolution (O(few~$\mu$m)) represent a 
new avenue in the development of silicon trackers, enabling a variety of physics studies which would remain out of experimental 
reach with the existing technologies. Several technology solutions are being currently pursued by the community to address the 
challenges posed by various experiments~\cite{arxiv.2203.13900,arxiv.2203.06773,arxiv.2203.08554}, both for the 
sensors~\cite{arxiv.2202.11828} and the associated electronics. The packaging and integration of the sensors and readout 
electronics will become more critical for future experiments, as device segmentation decreases to mitigate the increased track 
density. Bump bonding technologies have nearly reached their limits; more advanced 3D-packing technologies including wafer-to-wafer 
and die-to-wafer hybrid bonding and through silicon vias (TSV) have the potential to meet the goals of future particle physics experiments~\cite{arxiv.2203.06093,arxiv.2203.07626}.

Detector mechanics will also play a significant role in future detectors' performance. Material necessary for cooling and structural 
stability will be the lower bound on the radiation length for future tracking systems. Increased segmentation will lead naturally to 
larger power densities; in order to minimize material, solutions with integrated services and cooling are necessary. A holistic 
approach to design, simulation and manufacturing will be required. Novel materials, new cooling and composite manufacturing technique 
will need to be developed in order to reach the targeted performance~\cite{arxiv.2203.14347}. 

To develop these new technologies, simulations of the properties of silicon and novel sensor materials throughout the lifetime of 
the experiments will be critical. These studies can drive device design included implant locations, size and strength to the most 
promising directions of development~\cite{arxiv.2203.06216}. To reach their full potential, further developments are needed to 
improve accuracy, precision and new devices. With this research, the performance of future experiments can be better predicted for 
its full life cycle prior to construction.

\subsection{Simulations of Silicon Radiation Detectors for HEP Experiments}

There are currently a variety of tools available for simulating the properties of silicon sensors before and after irradiation. These 
tools include finite element methods for device properties, dedicated annealing models, and testbeam and full detector system models. 
No one model can describe all of the necessary physics. Most of these models are either fully or partially developed by HEP scientists 
and while there are many open-source tools, the most precise device property simulations rely on expensive, proprietary software. 

While existing approaches are able to describe many aspects of signal formation in silicon devices, even after irradiation and 
annealing, there is significant R\&D required to improve the accuracy and precision of these models and to be able to handle new 
devices (e.g. for timing) and the extreme fluences of future colliders. The US particle physics community can play a key role in this 
critical R\&D program, but it will require resources for training, software, testbeam, and personnel. 

For example, there is a great need for (i) a unified microscopic model of sensor charge collection, radiation damage, and annealing (no 
model can currently do all three), (ii) radiation damage models (for leakage current, depletion voltage, charge collection) with 
uncertainties (and a database of such models), and (iii) a measurement program to determine damage factors and uncertainties for particle 
types and energies relevant for current and future colliders. Equipped with these tools, the US community can play a key role in the 
success of future collider physics, and help in defining the most promising avenues to realize the necessary performance of future 
experiments. 

\subsection{Novel Sensors for Particle Tracking}

The proposed future collider experiments pose unprecedented challenges for event reconstruction, which require development of new 
approaches in detector designs. Several technologies are currently being investigated to develop technologies for the future, which 
approach these goals in complementary ways. 

Silicon or diamond sensors with 3D-technology have electrodes oriented perpendicular to the plane of the sensor. New 3D-geometries 
involving p-type trench electrodes spanning the entire thickness of the detector, separated by lines of segmented n-type electrodes for 
readout, promise improved uniformity, timing resolution, and radiation resistance. Present research aims for operation with adequate 
signal-to-noise ratio at fluences approaching $10^{18}~n_{\rm eq}/{\rm cm}^2$, with timing resolution on the order of 10 ps. 

Monolithic Active Pixel Sensors (MAPS), in which charge collection and readout circuitry are combined in the same pixel, have been 
shown to be a promising technology for high-granularity and low material budget detector systems. MAPS have several advantages over 
traditional hybrid pixel detector technologies, as they can be inexpensively fabricated in standard CMOS imaging processes, 
back-thinned or back-illuminated, have demonstrated high radiation hardness. Many MAPS pixel geometries have been explored, but the 
close connection of a sensor and front-end amplifier, without the need for external interconnections, holds the promise of reducing 
the input capacitance significantly, and hence extremely low-noise designs are possible. The reduction of the noise floor means that 
even small signals, for example from thinned low-power sensors, can yield satisfactorily high signal-to-noise. 

The DoTPiX pixel architecture has been proposed on the principle of a single n-channel Metal-Oxide Semiconductor (MOS) transistor, 
in which a buried quantum well gate performs two functions: as a hole-collecting electrode and as a channel current modulation gate. 
The quantum well gate is made with a germanium silicon substrate. The active layers are of the order of 5 $\mu$m below the surface, 
permitting detection of minimum ionizing particles. This technology is intended to achieve extremely small pitch size to enable 
trigger-free operation without multiple hits in a future linear collider, as well as simplified reconstruction of tracks with low 
transverse momentum near the interaction point. 

Furthermore, a technology is under development in which a novel ultra-fast scintillating material employs a semiconductor stopping 
medium with embedded quantum dots. The candidate material, demonstrating very high light yield and fast emission, is a gallium arsenide 
(GaAs) matrix with indium arsenide (InAs) quantum dots. The first prototype detectors have been produced, and pending research goals 
include demonstration of detection performance with minimum ionizing particles, corresponding to signals of about 4000 electron-hole 
pairs in a detector of 20$~\mu m$ thickness. A compatible electronics solution must also be developed. While the radiation tolerance 
of the device is not yet known, generally quantum dot media are among the most radiation hard semiconductor materials.  

Thin film detectors have the potential to be fully integrated, while achieving large area coverage and low power consumption with 
low dead material and low cost. Thin film transistor technology uses crystalline growth techniques to layer materials, such that 
monolithic detectors may be fabricated by combining layers of thin film detection material with layers of amplification electronics 
using vertical integration.

\subsection{4-Dimensional Trackers}
\label{sec:if03_4Dtrackers}

Future collider experiments call for development of tracking detectors with 10-30~ps timing resolution, in addition to excellent 
position resolution, i.e. 4D-trackers. Time resolution of the future 4D-tracker detector can be factorized into contributions from 
the sensor itself, and those from the readout electronics. The overall detector system should contain a sensor with short drift 
time, high signal to noise, limited thickness in the path of a minimum ionizing particle (MIP) to reduce the Landau fluctuations, 
and small time-to-digital converter (TDC) bin size. Several technologies to address these needs are being developed and are 
introduced in this section. 

Several modifications to Low Gain Avalanche Detectors (LGADs) technology have been proposed and demonstrated to make them suitable 
for tracking sensors with 100\% fill-factor, achieving excellent timing and position resolution. A key feature of analog-coupled 
``AC-LGADs'' is the signal sharing between electrodes, which can be used to obtain the simultaneous 30~ps and 5~$\mu$m resolutions. 
Another design geared towards 100\% fill factor are the ``Deep-Junction'' (DJ-LGAD), which are formed by abutting thin, 
highly-doped p$+$ and n$+$ layers. ``Buried Layer LGADs'' aim to achieve higher radiation tolerance by implanting the boron layer 
at low energy, and then burying it under a few microns of epitaxially grown silicon. 

Another thrust to develop 4D-tracking sensors is through a usage of novel sensors with 3D-geometries, or closer integration with 
electronics via monolithic pixel sensors, or adoption of new materials in sensor design. Going even beyond 4D-sensors are designs 
that aim to simultaneously measure not only the position and time, but also the angle of passing tracks. These kind of sensors 
would dramatically reduce the complexity of detector modules, and enable unprecedented reconstruction capabilities on the 
front-end. ``Double Sided LGADs'' (DS-LGADs) achieve this goals by adding a readout layer to the p-side of the LGAD structure, which 
allows one to also measure signals from the slower-drifting holes.

The timing Application Specific Integrated Circuits (ASICs) under development for the HL-LHC timing upgrades, named 
ALTIROC (ATLAS) and ETROC (CMS), represent revolutionary steps forward as the first readout chips to bring O(10~ps) timing to 
collider experiments. However, they are able to use significantly more space and power than high density ASICs designed for 
trackers with fine pitch and limited material. Compared to the RD53A ASIC designed for the HL-LHC pixel tracker upgrades, 
the timing chips use several hundred times more power and area per channel. The primary challenges to transform them into chips for 
4D-tracking will be to minimize both the power consumption and the channel size. Currently there are several projects with the 
aim to make advances in these areas. 

\subsection{Integration}

In the past years, HEP experiments have been mostly relying on bump bonding for high-density pixel sensor-to-ASIC connection. The 
bump bonding technology was proven to be reliable; however, it is known to have several limitations: it only works down to 
20-50~$\mu m$ of pitch and has yield issues for finer connections. Furthermore, the solder balls used for the connection increase 
the input capacitance to the amplifier and hence the noise. With bump bonding,the sensor or chip need side extensions to have 
external connections. In terms of mechanical properties the resulting connection is subject to heat stress since it involves 
different materials; the minimum thickness is also limited since both chip and sensor need a thick enough support wafers in order 
to meet alignment and bow requirements for the bump bonding process. 

The introduction of more advanced packaging can solve many of these issues, allowing for the improvement of performance, yield and 
processing. 3D-integration is a common widespread technology in industry, it allows tight packaging of sensor and readout chip. 
Furthermore it allows to stacks multiple chips in a single monolithic device. There are many technologies available for 
3D-integration, hybrid bonding is the most widely accepted. TSVs allow multiple planes to be stacked and connected with external 
connections without the need of extensions or silicon interposers.

Advanced packaging and wafer to wafer bonding would facilitate several applications both in HEP and outside of HEP. Therefore, 
electronics and sensor advanced packaging provide a variety of technologies that can meet the needs of future particle physics 
experiments. Combining these capabilities with silicon technologies developed for HEP, such as LGADs and active edge sensors, will 
allow the design of sophisticated detector systems that can meet the increasing challenges of next generation experiments. The 
collaboration between research groups and industry with established expertise in advanced packaging is crucial for the successful 
introduction of this technology in the research community. 

Current availability of such technologies is mixed. Leading edge foundries and nodes now include 3D-processes (Intel Forveros, 
TSMC 3DFabric, Samsung x-cube) but are typically too expensive for HEP. Hybrid bonding is available from many vendors that already 
work with HEP. The availability of TSVs is more limited. Small pitch TSVs are available from some foundries at advanced nodes 
($>$32 nm), in Silicon-on-Insulator (SOI) wafers, and from specialty foundries. Larger pitch TSVs inserted in a completed wafer 
(via-last) are available from a variety of packaging suppliers. Reliable 3D-devices with these large pitch ($>$20 $\mu m$) 
external TSV connections can be accessed now with resources available to HEP. It is important to build design and fabrication 
experience with moderate scale companies now gearing up to address opportunities afforded by the US semiconductor initiative. 

\subsection{Mechanics}

Detector mechanics will play a significant role in future detectors' performance; the necessary improvements will require simulations, 
novel ways to reduce the total mass, as well as more integrated design concepts to save on material budgets and optimize performance. 
The increased segmentation has naturally lead to larger power densities requiring high performance material support structures with 
integrated services. In many cases, the material in these structures can be the limiting factor for the tracking performance of the 
system. Particle detectors at future colliders rely on ever more precise charged particle tracking devices, which are supported by 
structures manufactured from composite materials. Various engineering techniques able to solve challenges related to the design and 
manufacturing of future support structures have been developed.

Future particle colliders, such as the HL-LHC, the Future Circular Collider (FCC-ee, FCC-hh), 
or a muon collider will collide particles at unprecedented rates and present a harsh environment for future detectors. A 
holistic approach to the design and manufacturing of detector support structures will be necessary to achieve minimal weight systems. 
Novel techniques, materials and other design and manufacturing solutions provide an avenue to solve challenges of increasingly more 
complex and large tracking detectors. Complex stresses in composite structures pose a severe challenge to current simulation tools. 
Research and development efforts are underway to solve these challenges by exploring multi-functional composite structures.

Material savings due to novel approaches have the potential of reduction on the order of 30-50\% depending on more detailed R\&D 
studies. This is an ideal opportunity to explore the conjunction of latest techniques in composite engineering involving machine 
learning based algorithms for heat transfer, mechanical loading and micro-to-macro scale material response predictions. Different 
structures are being studied to achieve these goals which include: carbon fiber support structures with integrated titanium pipes 
for $\rm{CO_2}$ cooling, etched silicon and peak cooling micro-channels, Kapton-based support structures and engineered air cooling. 

\subsection{Collaborative Development Programs}

Collaborative efforts should be encouraged when possible to make efficient use of resources and reduce the financial burden on 
individual groups, and to enlarge the pool of expertise in the field. Especially critical is to encourage and support young 
researchers to engage and pursue intensive research programs to develop these new technologies, which will help to ensure an expert 
workforce is sustained for the long term. Considering the prohibitively high costs for productions of new technologies at commercial 
foundries, collaborative submissions can provide the only viable solution for research groups to access advanced production 
facilities. Such collaborations promise to significantly expand the possibilities for small experiments or small groups to join 
advanced detector R\&D programs, work with others in the field to build technology demonstrators, and help advance the solid-state 
tracking technology. 

\section{Trigger and Data Acquisition (DAQ)}
A trend for future HEP experiments is an increase in the data bandwidth produced from the detectors. Datasets of the 
Petabyte scale have already become the norm, and the requirements of future experiments -- greater in size, exposure, and complexity 
-- will further push the limits of data acquisition technologies to data rates of exabytes per seconds. The challenge for these future 
data-intensive physics facilities lies in the reduction of the flow of data through a combination of sophisticated event selection in 
the form of high-performance triggers and improved data representation through compression and calculation of high-level quantities. 
These tasks must be performed with low-latency (\textit{i.e.} in real-time) and often in extreme environments including high radiation, 
high magnetic fields, and cryogenic temperatures. The Snowmass Instrumentation Frontier Topical Group discussions have been summarized 
in~\cite{https://doi.org/10.48550/arxiv.2209.03794}.

Developing the trigger and data acquisition (TDAQ) systems needed by future experiments will rely on innovations in key areas:
\begin{itemize}
 \setlength{\itemindent}{1.5em}
\item[\bf IF04-1] Pursue innovations in the application of Machine Learning (ML) to TDAQ systems, particularly in the co-design of 
hardware and software to apply ML algorithms to real-time hardware and in other novel uses to improve the operational efficiency and 
sensitivity to new physics of future experiments;
\item[\bf IF04-2] Invest in the design of TDAQ system architectures that leverage new technologies, techniques, and partnerships to 
enable more intelligent aggregation, reduction, and streaming of data from detectors to higher-level trigger systems and offline data 
processing; and, 
\item[\bf IF04-3] Develop improved readout technologies that increase data bandwidth and are capable of operating in extreme 
environments, while fitting the material and power constraints of future experiments.
\end{itemize} 

Critically, innovations in TDAQ rely on the people and processes behind them, and require investments in those people and 
infrastructure for R\&D. To that end, we call for:
\begin{itemize}
 \setlength{\itemindent}{1.5em}
\item[\bf IF04-4] Increased effort to build and retain domain knowledge for complex TDAQ systems by reliably supporting facilities 
and people – particularly engineers and technical staff, and early-career scientists through recruitment and training – in order to 
bring new ideas from early design and prototyping all the way through integration, commissioning, and operation in future detectors; 
and,
\item[\bf IF04-5] The creation of a dedicated (distributed) R\&D facility that can be used to emulate detectors and TDAQ systems, 
offer opportunities for integration testing (including low- and high-level triggering, data readout, data aggregation and reduction, 
networking, and storage), and develop and maintain an accessible knowledge-base that crosses experiment-project boundaries.
\end{itemize}

\subsection{A Compelling Use Case: Track-based Triggers}

Searches for novel new physics are often made possible by the development of novel new triggers that take advantage of improvements 
in detectors and real-time computing. Early-stage trigger systems of experiments at hadron collider experiments, for example those 
at the Large Hadron Collider (LHC~\cite{LHC}), often select objects with high momentum, based on calorimeter information, in order to reduce 
the large backgrounds to events of physical interest. However, this process necessarily loses many events, and is poorly optimized 
to explore a variety of potential beyond Standard Model (SM) scenarios. For instance, low momentum events with displaced vertices 
could lead to soft-unclustered-energy-patterns, long-lived staus or the decays of long-lived dark scalars in the Higgs portal 
scenario~\cite{wp_exotracktrig}. Alternatively, high-momentum and short-lived particles, such as particle dark matter, might be 
missed by the existing trigger systems due to their invisible decay within the detector volume~\cite{wp_tracktrig}.

For both of these cases, early-stage triggers based on tracking information could preserve interesting signal events for study. A 
difficulty of including tracking information in early-stage triggers is the complexity of calculating tracks and the speed at which 
decisions need to be made. For instance, at the LHC events are produced at a rate of 40 MHz, and this must be very 
rapidly reduced to the kHz level at the first step of the trigger system. The development of a fast track-based trigger is therefore 
an area of considerable interest in the HEP community, and several different approaches have been taken to solve the 
problem for silicon-strip tracking systems. For future hadron collider experiments, a fast tracking trigger for silicon pixel 
tracking layers at small radii with respect to the beam pipe, and therefore with significantly more channels than strip trackers, 
would open the sensitivity to particles arising from beyond SM physics in an interesting lifetime regime.

As a new approach, a white paper contributed to Snowmass~\cite{wp_tracktrig} utilizes highly-parallelized graph computing architecture 
using field-programmable gate arrays (FPGAs) to quickly performing tracking in small-radius silicon detectors. This paper proposes 
unsupervised machine learning on a highly-parallelized graph computer constructed using modern FPGA technology. As documented 
in~\cite{GraphTracking}, this algorithm yields a trigger efficiency near 100\% for track transverse momenta above 10~GeV and a 
spurious trigger rate of a few kHz. Implementation studies on a Virtex Ultrascale+ FPGA are in progress and indicate a latency of 
200~ns, well below the 4 $\mu$s limit for trigger algorithms at LHC experiments.

We note that the design of future detector systems, especially tracking systems in dense, high occupancy environments, is best done 
taking into account the necessary trigger and DAQ considerations early in the process. For example, the design of the HL-LHC silicon 
tracker for the CMS experiment~\cite{CMSPhase2Tracker} utilizes the local coincidence of hits in two closely-spaced tracking layers 
to provide some limited momentum selection capability. This is necessary to reduce the data rate from the detector to manageable 
levels for the early-level trigger system. 

\subsection{Heterogeneous Computing and Machine Learning}

As is well known, the scaling of single Central Processing Unit (CPU) solutions to efficiently address computational problems has 
reached its limits, and the need for the parallelization of tasks across many CPU cores, in Graphical Processing Units (GPUs), or 
in other specialized hardware like FPGAs and ASICs has become necessary. The choice of technology often depends on the timescale 
and amount of data to handle. For high-energy physics experiments this means moving beyond the natural parallelization of processing 
individual data ``events'' (e.g. beam crossings, which are independent of each other) to the parallelization of the algorithms 
themselves acting upon the data of single event. This requires a new paradigm of coding algorithms to take advantage of the available 
heterogeneous computing hardware, and thus utilizing industry tools or developing domain specific tools to aid in this parallelization. 

ML algorithms lend themselves well to be distributed on such heterogeneous computing platforms using standard libraries, 
and thus make a natural and powerful target for trigger applications. The applications include the very specialized and local 
processing at the front-end of the detector electronics (``edge'' computing) where low-level detector hits are converted into clusters 
or other higher-level data objects at high frequency and low latency, but also the more global and generalized processing needed to 
discriminate physics signatures from backgrounds. ML also could potentially be used to go beyond the fixed hand-curated 
trigger menus used to select physics data at colliders to a novel ``self-driving'' paradigm whereby the trigger system autonomously 
and continuously learns from the data to more efficiently and effectively filters and selects data from a detector system, as 
discussed in white paper~\cite{TDAQ-innovations}. Such systems may complement dedicated triggers, searching for specific signatures of 
Beyond-Standard-Model physics, by performing a general set of anomaly detection that may be sensitive to a wider variety of new physics.

Development of ML algorithms for use in the trigger and data acquisition systems of future HEP experiments 
is particularly challenging~\cite{fastml_summary}. For example, while ML-based algorithms have proven effective at performing  
data reduction, through both advanced data selection and data compression, to be used in high-luminosity environments of future particle 
colliders these algorithms must be capable of running in on-detector electronics with latencies on the order of nanoseconds. Future 
neutrino, DM, and astrophysics experiments require latencies on the order of microseconds to milliseconds, and often 
similarly extreme detector environments.

The design and development of low-latency artificial intelligence (AI) algorithms requires optimization across both physics 
(\textit{e.g.} selection efficiency, background rejection) and technical performance (\textit{e.g.} latency, resource usage). 
It's necessary to consider a co-design of hardware and software that gives special attention to the processor element, making 
use of tools and expertise that can bring a variety of ML algorithms built on large datasets into FPGA and ASIC firmware. 
Open-source frameworks like \texttt{hls4ml}~\cite{hls4ml} and FINN~\cite{FINN} aim to ease the complexity of firmware programming, 
which have opened up development and integration of sophisticated AI into high-performance hardware. Continued development of ML 
frameworks that can aid hardware-software co-design, coupled with (and in many cases driving) improvements in the underlying 
processor technologies, can open the door to paradigm shift in how future experiments will collect, reduce, and process data.

\subsection{Innovative Architectures}

For experiments where a processing element of the trigger system must have a complete view of the data from all detectors to perform 
its function, scalability of the trigger system can be achieved by time multiplexing the data to individual processors. This is 
natural for software-based trigger levels (CPUs, GPUs) in collider experiments, where data are aggregated in an event builder and sent 
to a target compute node for processing asynchronously. For a hardware-based synchronous trigger level (FPGAs, ASICs), this can be 
achieved by sending data from all detector elements for a given time slice (or event) via links to target processors in a round-robin 
fashion (``time multiplexing''). This will be necessary for the CRES tritium beta decay experiment, for example, where each compute 
node must process the data from all receivers in a given time slice.

In collider experiments, trigger data processing in the hardware-based level is still generally synchronous to the accelerator clock, 
even if the processing is time multiplexed. At all stages the data are processed and registered at a multiple of this frequency in the
digital pipeline, and the event number is implicit by the clock (or accelerator bunch) counter. However, it is enough to time-stamp 
data at the very front end of the detectors with the system clock, and transmit and process the data asynchronously as traditionally 
done at the software-based level~\cite{async-L1}. Effectively the event builder infrastructure moves to the first level of the trigger
system. This would alleviate the challenge of distributing and synchronizing a stable, low-jitter, high frequency clock over the
entirety of a very large and distributed electronics system. It also has the additional benefit of blurring the lines between the
first level trigger, which typically runs in fast FPGAs, with the software-based higher levels. Another avenue of potential interest 
in this area is to make use of neuromorphic computing, i.e. the use of very-large-scale integration systems containing electronic 
analog circuits to mimic neuro-biological architectures present in the nervous system, directly on analog signals. 

Another innovative approach to solving the data reduction problem for trigger and data acquisition systems is to move away from a 
pipelined and triggered readout, and instead operate in a more ``streaming" design, where data is encoded with its time and 
origin~\cite{TDAQ-innovations}. In this model, pioneered by the LHCb experiment for its upgrade but being adopted to some degree as 
well by the other LHC experiments, event data can be reduced at its source, often through simple thresholding and zero suppression, 
and then aggregated and streamed to downstream computational and storage elements. There, full- or partial-event filtering and 
further processing and translation of data into higher-level quantities can be performed in order to achieve the reduction in the data 
throughput and offline computing. Hybrid designs that combine both traditional trigger-based DAQ for some detector subsystems and 
streaming-readout for others is also possible. Emphasis on such approaches for upgrades and experiments at future facilities has 
merit, especially due to its ability to simplify DAQ design.

\subsection{Developments in Novel Readout Technologies}

The needs of future detectors continue to push readout, triggering, and data acquisition technologies to operate with growing data 
rates in more extreme environments. Future kton-scale neutrino and dark matter experiments like DUNE and LZ will produce many 
petabytes of data per year with intrinsic data rates in excess of TB/s, and require readout systems that can reliably operate in 
cryogenic temperatures over the long lifetimes of the experiments and minimize radiological material volumes to maintain sensitivity 
to low-energy interactions. In high-energy collider physics at the HL-LHC or potential future colliders like the FCC-hh, data rates 
in the hundreds of TB/s are possible from tracking and calorimetry systems, and must be able to withstand high radiation rates and 
not significantly add to the material budget of the detectors. The fundamental challenge of how to move data from readout 
electronics to online and offline computing resources requires a commitment to research and development in new and improved 
technologies.

Core to improvements in DAQ are a combination of reducing the data rate close to the detector, and increasing the data bandwidth for 
a given material and/or power cost, in the extreme environments required. More sophisticated data reduction techniques in detector 
electronics may be possible with advances in AI, particularly with improvements in translating low-latency machine-learning-developed 
compression and triggering algorithms to ASICs~\cite{fastml_summary}. In many cases, like in fast tracking algorithms, correlations 
across different portions of the detector are necessary to develop effective trigger algorithms, and thus fast, localized, and 
low-material communication is necessary. Wireless communication technologies are an area of large promise here: microwave-based 
technologies already show reliable data transmission at the 5 Gb/s scale, and promising future work using free space optics may 
allow for wireless communication at the Tb/s scale~\cite{readout-innovations}. Integration of wireless communication into HEP 
detector design (like, for future tracking detectors in colliders), could allow for new system designs that exploit localized 
readout, fast analysis, and triggering to intelligently reduce data volumes.

Technologies to allow greater bandwidth off of the detector also show significant promise~\cite{readout-innovations,readout-links}. 
Silicon-photonics are an appealing alternative to the current Vertical Cavity Surface-Emitting Laser (VCSEL)-based approaches: 
they can allow integration of fiber-optic connections directly to sensor modules or readout chips (thus reducing the need for 
electrical cable connections), commercial devices already show high radiation tolerance, and offer a bandwidth twice that of VCSEL  
devices with a power consumption that is 20\% less. More promising developments exist in Wavelength Division Multiplexing (WDM), 
where individual serial links can be transmitted on its own wavelength, reducing the need for data aggregation to maintain high 
data bandwidth per link. WDM could be used in a design to bring data out of the most extreme radiation environments in colliders 
more efficiently, allowing for further data reduction in downstream DAQ components.

Notably, the work to develop new readout technologies is often less in design, but more in integration and testing with real 
detector components in real detector environments. It is important to develop and maintain tools and facilities that can allow 
for realistic full-system testing of readout electronics and DAQ. 

\subsection{Timing}

Time measurements will feature prominently in the next generation of particle physics experiments and upgrades, including 
integration into 4D-tracking systems and 5D-calorimeters (see~\ref{sec:calorimetry-timing}). Timing helps disentangle the effects of
particle pile-up in hadron collisions, discriminates against beam-induced backgrounds in a muon collider experiment, and can be
used to separate Cherenkov and scintillation light signals in neutrino and dark matter experiments. Timing also provides particle 
ID information useful for a broad range of experiments, including sensitivity to any slow beyond-standard-model long-lived
particles. Thus it seems evident that timing information will work its way into the trigger processing chain to improve its 
selectivity and precision of measurements. A particular challenge is the need for synchronization of data at the ${\cal O}(10)$~ps 
level or better, including across large distances in the case of RF arrays. 

\subsection{Fostering and Retaining Expertise in TDAQ}

Along with the importance in conducting the R\&D to develop and ultimately construct innovative trigger and data acquisition 
systems for future physics facilities, equally important is to build and retain the domain knowledge and technical expertise 
within the high-energy physics community required to support this~\cite{TDAQ-innovations}. It can be challenging to recruit and
retain highly skilled personnel to address the specialized and high-tech needs of our community. Essential technical staff can 
leave for higher-paying positions in industry, and younger scientists specializing in instrumentation may find career 
progression and promotion a challenge in this field. This is compounded by the timescale for large experiment facilities from 
construction through to the end of operations, which can be decades. Thus it is imperative for the scientific community to provide 
career opportunities for such highly skilled people.

To facilitate fast machine learning development, it is very important to foster interdisciplinary collaboration between electrical 
engineering, computer science and physics, as people in these fields have valuable expertise in digital design, machine learning 
techniques and the physical problems to be addressed. Also, as much as possible, work should be preserved in an open source
manner, to be used cross-project and cross-experiment and further built upon in the future.

However, the needs of particle physics experiments are not always relevant to industrial partners, which implies that the retention 
of highly-skilled personnel is made even more important. The particle physics community has an interest in edge cases to typical 
uses in industry, such as electronics that perform well under constant bombardment from radiation or in cryogenic fluids. There is 
also a need for a high degree of reliability, as many devices are installed in detectors in areas that are inaccessible for
replacement for decades at a time. It is important for the community to continue to follow development in industry and to build strong
collaborative networks, but it is as important to develop resources to pursue R\&D directions that are specific to use cases in the 
particle physics domain. 

As we explore new TDAQ architectures and hardware with increasing complexity whose performance depends on interactions on a systems
level, a dedicated facility that can support TDAQ design and development while also offering opportunities for integration testing
across low- and high-level triggering, data readout, data aggregation and reduction, networking, and storage should be established. 
While the necessary support hardware for the facility may be localized, the participating domain experts should encompass a 
distributed community. Given many of the common challenges across physics frontiers, such a Trigger and Data Acquisition Emulation 
and Integration Test Facility should cross experiment and project boundaries, offering support for emulation of detectors and TDAQ
systems, and the development and maintenance of common hardware, firmware, and software, and to support TDAQ R\&D for future
detectors. This facility will develop and maintain an accessible knowledge-base by enabling and supporting connections and
communication between engineers and scientists in many National Labs and university labs working within different subfields that 
encounter similar problems.

Finally, we note that the success of a detector upgrade or new experiment ultimately depends not just on the delivery of the new
components, but on the successful installation, commissioning, and integration of them into the experiment as well as their 
efficient operation. This applies to trigger-DAQ as much as any other instrumentation area. These tasks need to be well thought out
(preferably as part of the initial proposal to evaluate the overall cost of a new system) and supported. Lack of attention in any 
of these areas can lead to substantial and costly delays as well as a failure to reach the design goals, which jeopardizes the 
physics output. Thus the installation, commissioning, and integration tasks as well as the long-term operations (until the end of 
the experiment) must be a priority.

\section{Micro-Pattern Gaseous Detectors}

Gaseous Detectors are the primary choice for cost effective instrumentation of large areas and for continuous tracking of charged 
particles with minimal detector material. Traditional gaseous detectors such as the wire chamber, Resistive Plate Chamber (RPC), and 
TPC with multiwire proportional chamber (MWPC) readout remain critically important for muon detection, track-finding, and triggering 
in ongoing and planned major particle physics experiment, including all major LHC experiments (ALICE, ATLAS, CMS, LHCb) and DUNE.

Micro Pattern Gaseous Detectors (MPGDs) are gas avalanche devices with order $\mathcal{O}$(100~\textmu m) feature size, enabled by 
the advent of modern photolithographic techniques. Current MPGD technologies include the Gas Electron Multiplier (GEM), the Micro-Mesh 
Gaseous Structure (MicroMegas), Thick GEMs (THGEMs), also referred to as Large Electron Multipliers (LEMs), the Resistive Plate WELL 
(RPWELL), the GEM-derived architecture ($\mu$RWELL), the Micro-Pixel Gas Chamber ($\mu$-PIC), and the integrated pixel readout 
(InGrid).

MPGDs have already significantly improved the segmentation and rate capability of gaseous detectors, extending stable operation to 
significantly harsher radiation environments, improving spatial and timing performance, and even enabling entirely new detector 
configurations and use cases.

In recent years, there has therefore been a surge in the use of MPGDs in nuclear and particle physics. MPGDs are already in use for 
upgrades of the LHC experiments and are in development for future facilities, such as the Electron Ion Collider (EIC), the 
International Linear Collider (ILC), FCC, and the Facility for Antiproton and Ion Research (FAIR). More generally, MPGDs 
are exceptionally broadly applicable in particle physics, hadron physics, heavy-ion physics and nuclear physics, charged particle 
tracking, photon detectors and calorimetry, neutron detection and beam diagnostics, neutrino physics, and dark matter detection, 
including operation at cryogenic temperatures. Beyond fundamental research, MPGDs are in use and considered for scientific, social, 
and industrial purposes; this includes the fields of material sciences, medical imaging, hadron therapy systems, and homeland security. 

Five commissioned white papers on MPGDs were developed during the 2021 Snowmass decadal survey. These summarize ongoing R\&D on 
MPGDs~\cite{Dehmelt:2022inw}, the future needs for MPGDs in nuclear physics~\cite{Barbosa:2022zql}, and future needs for MPGDs in 
three broad areas of particle physics: low-energy recoil imaging~\cite{OHare:2022jnx}, TPC readout for tracking at lepton 
colliders~\cite{Bellerive:2022wrb}, and tracking and muon detection at hadron colliders~\cite{Black:2022sqi}. A white paper with 
further details on a proposed TPC tracker for Belle-II was also submitted~\cite{Centeno:2022syq}. These whitepapers are summarized 
in our Snowmass Topical Group report on MPGDs~\cite{Surrow:2022ptn}. An abbreviated version of this summary follows below.

\begin{itemize}
 \setlength{\itemindent}{1.5em}
\item [\bf IF05-1] MGPDs constitute an enabling technology that is key for large segments of the future US Nuclear Physics (NP) 
and HEP programs, and which also benefits other communities. MPGDs provide a flexible go-to solution whenever particle detection 
with large area coverage, fine segmentation, and good timing is required.
\item [\bf IF05-2] The technology is relatively young and should be advanced to performance limits to enable future HEP experiments. 
Support of generic and blue-sky R\&D is required to achieve this.
\item [\bf IF05-3] The global HEP community would benefit from US strategy coordination with the ECFA detector R\&D implementation 
process in Europe.
\item [\bf IF05-4] In order to maintain and expand US expertise on MPGDs, The US NP and HEP communities would benefit strongly 
from a joint MPGD development and prototyping facility in the US.
\end{itemize}

\subsection{Recent Advances, Current R\&D, and Future Needs}

Recent developments in the field of MPGDs, and the role of the RD51 collaboration, a CERN research collaboration for the 
development of MPGD detectors, are summarized in Ref.~\cite{Dehmelt:2022inw}. MPGDs were developed to cost-effectively cover 
large areas while offering excellent position and timing resolution, and the ability to operate at high incident particle rates. 
Significant development time was invested in optimizing manufacturing techniques for MPGDs, in understanding their operation, and 
in mitigating undesirable effects such as discharges and ion backflow. The early MPGD developments culminated in the formation of 
the RD51 collaboration hosted by CERN, which has become the critical organization for promotion of MPGDs and which coordinates all 
aspects of their production, characterization, simulation and use in an expanding array of experimental configurations. The CERN 
MPGD Workshop is a source of essential expertise in production methods, mitigation and correction of manufacturing issues, and the 
development of MPGDs for specific experimental environments. 

An impressive array of MPGDs has been developed, from the initial GEM and MicroMegas, now used in a wide variety of applications and 
configurations, through the more recent THGEMs, and $\mu$RWELLs with resistive layers to mitigate discharge effects. MPGDs are 
now also used jointly with other detector elements, for example with optical readout and in LAr detectors. In parallel with 
MPGD detector development, there has been an important creation of a standardized, general electronics system, the Scalable Readout 
System (SRS). This system has seen widespread use and is of great utility in allowing integration of a variety of frontend (FE) 
electronics into one data acquisition system. For Snowmass 2021, a number of Letters of Interest were received that illustrate 
ongoing developments and expansion of use of MPGDs. Submissions on high-precision timing, high-rate applications, expansion of 
the SRS readout system triggering capabilities, and reduction of ion backflow are summarized in Ref.~\cite{Dehmelt:2022inw}. 
Many other improvements of MPGDs are being actively pursued. Several of these efforts are commented on below. A detailed, 
comprehensive summary of expected MPGD needs of planned HEP experiments versus time can be found in the 2021 ECFA detector 
research and development roadmap~\cite{Detector:2784893}, specifically in Fig.~1.1.

\subsection{MPGDs for NP Experiments}
Many current and future NP experiments in the US have or are implementing MPGDs for tracking and particle identification (PID) 
purposes. We have summarized the role that MPGDs play in NP experiments, and the R\&D needed to meet the requirements of future 
NP experiments in Ref.~\cite{Barbosa:2022zql}. Examples include advanced MPGDs for Tracking at the EIC, to be built at 
Brookhaven National Laboratory (BNL), which requires its tracking system to have low mass (X/X$_0 \leq$ 1\%), large area 
$\mathcal{O}$(1~m$^2$), and excellent spatial resolution $\mathcal{O}$(100~\textmu m). MPGDs such as the GEM, MicroMegas, 
and $\mu$RWELL can meet these requirements. The Facility for Rare Isotope Beams (FRIB) at Michigan State University will become 
the world's most advanced facility for the production of rare isotope beams (RIBs). MPGD technologies play an essential role in the 
science program's success at FRIB. Applications of MPGD technologies include low-pressure tracking and PID at the focal planes of 
magnetic spectrometers, Active-Target TPCs, and TPCs for the detection of exotic decay modes with stopped RIBs. Future 
spectrometers for NP experiments at the Thomas Jefferson National Accelerator Facility (JLab) require large area 
$\mathcal{O}$(m$^2$), low mass (X/X$_0 \leq$ 1\%), excellent spatial resolution $\mathcal{O}(100\, $\textmu m), excellent timing 
$\mathcal{O}$(10 ns), high rate $\mathcal{O}$(1~MHz/cm$^2$) tracking detectors for operation in high background rate and high 
radiation environment. Only MPGD technologies such as GEMs, MicroMegas, or $\mu$RWELL detectors can satisfy the challenges of high 
performances for large acceptance at reasonably low cost. 

{\bf Dedicated MPGD Development Facility} 
Currently, the majority of MPGD developers and users in the US rely on production facilities and expertise, diagnostic facilities, 
and standardized readout electronics associated with the RD51 collaboration and CERN. This significantly slows down the R\&D cycle 
and limits the speed of innovation in the US. It also means all production for US-led experiments has to be outsourced. A US-based 
MPGD Center of Excellence is needed and would address this issue. We envision a facility similar in nature to the Gaseous Detector 
Development (GDD) lab at CERN or the Silicon Detector facility at the Fermi National Accelerator Laboratory (FNAL). Such a facility 
would benefit both the nuclear physics and particle physics communities in the US. There are also ample opportunities for 
commercialization and collaboration with industry. We envision such a facility hosted by one of the Department of Energy's (DOE) 
National Labs, such as JLab or BNL.

\subsection{MPGDs for Recoil Imaging in Directional Dark Matter and Neutrino Experiments}
MPGDs can be used to read out the ionization in low-density gas TPCs with exquisite sensitivity and spatial resolution. In the most 
advanced MPGD TPCs, even individual primary electrons -- corresponding to an energy deposit on the order of 
$\sim$30~eV -- can be detected with negligible background from noise hits, and 3D-ionization density can be imaged with 
$\sim$(100~\textmu m)$^3$ voxel size. This new experimental technique, enabled by MPGDs, has a large number of interesting 
applications in fundamental and applied physics. 

One of the most intriguing applications is to scale up TPCs with MPGD readout to construct a competitive, low-background, 
high-definition ionization imaging experiment, CYGNUS, which would be directionally sensitive to both dark matter and neutrinos. There is 
an opportunity here for the US to take the lead and initiate a novel experimental program focused on recoil imaging, with broad scientific 
scope, one that we have only just started to map out. Because large areas, $\mathcal{O}$(1000~m$^2$), of MPGDs are required, there are 
clear synergies with this proposal and the needs in nuclear physics, where MPGDs will be used broadly, and where R\&D, production, and test 
facilities in the US are also desirable. Other notable applications of recoil imaging include the International Axion Observatory (IAXO), 
directional neutron detection, the measurement of the Migdal effect, X-ray polarimetry, the detection of rare nuclear decays. Several 
groups are also exploring TPC designs to use high density gases such as SeF$_6$ or argon and can provide the sub-mm to 10~micron resolution, 
for example via `dual readout' TPCs which can detect both the positive ions as well as the electrons generated by a recoil event. TPCs 
using gaseous argon could be of interest for studies of the neutrino sector, for example $\tau$-tracking for the study of $\nu_\tau \tau$ 
charged current interactions. Further detail on the emerging field of recoil imaging can be found in~\cite{OHare:2022jnx}.

\subsection{MPGDs at Future HEP Colliders}

Advances in our knowledge of the structure of matter during the past century were enabled by the successive generations of high energy 
particle accelerators, as well as by continued improvement in detector technologies. In this context, MPGDs have become a preferred 
solution for enabling both continuous, low-mass charge-particle tracking in TPCs and large-area muon detection systems. Two of the most 
prominent MPGD technologies, the GEM and MicroMegas, have been successfully operated in many different experiments, such as Compass, 
LHCb, and TOTEM. In addition, the low material budget and the flexibility of the base material makes MPGDs suitable for the development 
of very light, full cylindrical fine tracking inner trackers at lepton colliders such as KLOE-2 at the Double Annular $\Phi$ Factory 
for Nice Experiments (DAFNE) and BESIII at the Beijing Electron–Positron Collider II (BEPCII).

The TPC concept is viewed in particle physics as the ultimate drift chamber since it provides 3D-precision tracking with low material 
budget and enables particle identification through dE/dx measurements with cluster counting techniques. At ILC and the Chinese 
Electron Positron Collider (CepC), as well as for Belle-II upgrades, TPCs with MPGD readout are key to proposed detector designs~\cite{Bellerive:2022wrb,Black:2022sqi,Centeno:2022syq}.

Gaseous detectors are also the primary choice for cost effective instrumentation of very large areas, with high detection efficiency in 
a high background and hostile radiation environment, needed for muon triggering and tracking at future facilities. They can provide a 
precise standalone momentum measurement or be combined with inner detector tracks resulting in even greater precision. Adding precise 
timing information, $\mathcal{O}$(ns), allows control of uncorrelated background, mitigates pile-up and allows detection of extremely 
long lived particles that behave like slow muons propagating through the detector volume over a time as long as a few bunch crossings. 
Scaling up of MPGDs to very large single unit detectors of $\mathcal{O}$(m$^2$), has facilitated their use in muon systems in the LHC 
upgrades. Major developments in the MPGD technology have been introduced for the ATLAS and CMS muon system upgrades, towards 
establishing technology goals, and addressing engineering and integration challenges. MicroMegas and GEM have been recently installed 
in the ATLAS New Small Wheel and the CMS GE1/1 station respectively, for operation from LHC Run 3 onward, as precise tracking systems. 
Those radiation-hard detectors, able to cope with the expected increased particle rates, exhibit good spatial resolution, 
$\mathcal{O}$(100 \textmu m) and have a time resolution of 5--10~ns. In the CMS muon system additional stations, GE2/1 and ME0, based 
on GEMs with high granularity and spatial segmentation, will be installed to ensure efficient matching of muon stubs to offline pixel 
tracks at large pseudo-rapidities during HL-LHC operation. Several solutions ($\mu$RWELL, $\mu$-PIC, and small-pad resistive 
MicroMegas were also considered for the very forward muon tagger in the ATLAS HL-LHC Upgrade Muon Technical Design Report (TDR). Here, 
the main challenges are discharge protection and miniaturization of readout elements, which can profit from the ongoing developments on 
Diamond-Like Carbon (DLC) technology. The $\mu$RWELL is the baseline option for the HL-LHC Upgrade of the innermost regions of the 
Muon System of the LHCb experiment (beyond LHC Long Shutdown 4).

Muon systems at future lepton colliders, such as ILC, the CERN Linear Collider (CLIC), CepC, FCC-ee, and the Super Charm Tau Factory 
(SCTF) or the Large Hadron Electron Collider (LHeC), do not pose significant challenges in terms of particle fluxes and the radiation 
environment. Therefore many existing MPGD technologies are suitable for building future large muon detection systems. On the other 
hand, the expected particle rates for the muon tracking and triggering at future hadron colliders, such as the FCC-hh, make the 
existing technologies adequate in most regions of the spectrometers, but require a major R\&D for the very forward endcap region. In a 
multi-TeV muon collider, the effect of the background induced by the muon beam decays is extremely important. A new generation Fast 
Timing MPGD (FTM, Picosec) is considered to mitigate the beam induced background, by rejecting hits uncorrelated in time. 
\textbf{Required R\&D} should focus on stable operation of large area coverage, including precision timing information to ensure the 
correct track-event association, and on the ability to cope with large particle fluxes, while guaranteeing detector longevity using 
environmentally friendly gas mixtures and optimized gas consumption (gas recirculating and recuperation system. 

\section{Calorimetry}
\label{sec:calorimetry}

The Snowmass IF06 Calorimetry Topical Group has considered major issues in present and future calorimetry. Input has been taken from a series 
of talks, group discussions, Letters of Intent (LOI), and White Papers. Here we report on two major approaches to calorimeter systems 
-- Particle Flow (PF) and Dual Readout (DR), the critical extra dimension of precision timing, and the development of new materials 
for calorimeters. What follows is a distillation of the full report of the Topical Group on Calorimetry~\cite{White:2022abc}, 
highlighting the key points and recommendations. 

The primary purpose of EM and hadronic calorimetry is the measurement of the energy of charged and neutral particles and 
overall event energy. However, they are also important systems for overall event reconstruction, particle identification and 
triggering. The physics goals and the experimental conditions at future colliders require technical advances in calorimeter technology 
to fully exploit the physics potential of these facilities. For future $\ee$ colliders, so-called Higgs Factories, the overall 
precision of event reconstruction is the main focus, while future hadron colliders at energies and luminosities significantly beyond 
the HL-LHC impose new challenges in terms of the experimental environment.

The field of calorimeter research and development remains very active. The precision energy measurement requirements of future physics 
programs is stimulating innovation in the particle flow and dual readout systems. The increasing availability of precise timing is 
adding an important new dimension to system implementation, while the development of a range of fast, radiation-hard active materials 
is leading to increased flexibility and exciting possibilities for calorimeter system designs.

\begin{itemize}
 \setlength{\itemindent}{1.5em}
\item[\bf IF06-1] Future calorimetry at fixed target and colliding beam experiments should be \textbf{fundamentally multidimensional}, 
providing shower position, time, energy, and a detailed look at shower constituents through the exploitation of an application-specific 
\textbf{combination of PF techniques, materials with intrinsically good time or energy resolution, or DR techniques.} R\&D support is 
needed to achieve these goals.
\item[\bf IF06-2] Sustained R\&D is needed to move multidimensional calorimetry \textbf{from prototype to realistic detector.} Scaling 
to hundreds of thousands or tens of millions of channels while maintaining the required quality is a huge challenge.
\item[\bf IF06-3] \textbf{Electronics must be developed to allow new features} such as fast timing \textbf{without significantly adding 
to the power budget or cooling load.} It will be necessary to have \textbf{HEP personnel trained in the new technologies} and close 
tracking of new electronics options emerging in industry.
\item[\bf IF06-4] \textbf{Effective partnerships} with chemists, materials scientists, industries large and small, and radiation 
facilities must be continued and strengthened to \textbf{explore the landscape of materials} that enable precision calorimetry and to 
lower the cost.
\end{itemize}

\subsection{Precision Timing}
\label{sec:calorimetry-timing}

The potential for precision timing at the 10 ps level or better opens new possibilities for precise event reconstruction and the 
reduction of the negative effects of challenging experimental environments. Precise timing can directly benefit calorimetry in several 
ways ranging from detailed object reconstruction to the mitigation of confusion from pile-up. It can also lead to improved performance 
for both PF and DR-based calorimeters.

Given these possible performance enhancements, the focus is now on the study of timing implementation both at the device level and 
the calorimeter system level, together with detailed simulations. A range of possible implementations can be envisaged, addressing the 
needs of calorimetric measurements, confusion reduction, particle identification, and long-lived particle searches. The timing 
precision required varies across these applications. R\&D is needed to establish whether an average timing precision can satisfy all 
the requirements at some acceptable level, or, for instance, dedicated high precision timing components are needed in combination with 
a general systemwide implementation. 

The actual implementation of timing in calorimeter systems can take a number of forms. Timing information can be collected by all 
active calorimeter cells which, when implemented in a highly granular calorimeter enables a full 5-dimensional reconstruction of 
shower activity in the detector, with corresponding benefits for pattern recognition, spatial shower reconstruction and separation 
and energy measurement. However, the very large number of cells might force a compromise with only a fraction of cells instrumented 
for timing. A less challenging and expensive solution could be the use of timing layers, for instance before and after the 
EM calorimeter, and at certain depths in the hadronic calorimeter.

Careful studies are needed to assess the cost-to-benefits of possible system designs. Adding timing functionality at both the cell and 
layer levels has implications for material profiles/sampling ratios, power consumption, system complexity and expanded data volume. 
Due to recent developments a number of possible timing technologies have potential applicability to calorimeter systems. However, 
significant R\&D is needed to address the realities of these choices and this is seen as a major area of future calorimeter 
development. Successful implementation can lead to highly performant calorimeter systems well matched to the demands from both 
future physics studies and experimental environments.

\subsection{Active Materials}
\label{sec:calorimetry-materials}

The construction of future calorimeters matched to the demands of high radiation tolerance, the need for fast timing, and constrained 
cost puts strong requirements on the properties of active calorimeter materials. Suitable materials should therefore combine high 
density, fast decay time, and good radiation hardness with good optical quality and high light yield. To cope with unprecedented high 
event rate expected by future experiments, an ultra-fast total absorption calorimeter with sub-nanosecond decay time is preferred. A 
range of inorganic scintillators using rare-earth doping in crystals has been developed possessing many of these desired properties. 
However, use of these materials in future large-scale calorimeter systems demands attention to material costs to assure affordability 
and to radiation hardness for long attenuation length of scintillation light after high radiation dose. Scintillating glass is a new 
type of active material that is being engineered to have similar performance and radiation hardness to lead tungstate. Work continuing 
to scale up production to the sizes needed for large EM calorimeter in a cost-effective way is in progress. 

Organic-based scintillator calorimeters, including plastics scintillator, liquid scintillator, and water-based liquid scintillator, 
have fast pulse response with comparable light-yield to inorganic crystals. The plastic scintillators, polyvinyltoluene or 
polystyrene base, have seen very wide use, having the advantages of ease of use and relatively low cost. A new thermoplastics 
acrylic scintillator aiming to load scintillator materials and high-Z elements directly into acrylic monomers is under development. 
The combination of plastic scintillators with SiPM readout has significantly expanded the flexibility of calorimeter design. An 
alternative method to fabricate plastic detectors with tight controls on shape and dimension using a computer-aided 3D-printing design 
is highly desired. Finally, liquid scintillators have very low unit cost with improved properties in terms of stability and material 
compatibility that have been largely advanced through neutrino and other rare-event physics over the past decade and been widely used 
where very large volumes (tens of kilotons) are required. An interesting development is the potential of metal doped (water-based) 
liquid scintillators for use as active materials in sampling calorimeters.

\subsection{Particle Flow}
\label{sec:calorimetry-PF}

PF calorimetry makes use of the associations of charged tracks and calorimeter energy deposits to achieve precise 
reconstruction of hadronic jets and measurement of their energy. Such associations can also be effectively used to reduce the effects 
of pileup. Implementation of a Particle Flow Algorithm-based (PFA) calorimeter requires a small cell size and high granularity 
leading to very high channel counts. Challenges remain in realizing calorimeter systems with up to 100 million cells. Developent 
is onging for such systems in the areas of power and heat management, integration of on-board ASICs, and components of signal 
extraction. A variety of approaches to PFA calorimeters are under developent ranging from scintillator-SiPM systems to a range 
of gas-based systems using GEMs, MicroMegas and RPCs. Also being explored are the potential benefits of precise timing (see 
above) and moving some elements of a PFA into the front-end electronics.

The HEP community has demonstrated its strong interest in pursuing this technology in multiple applications in various stages of 
development. These include the HL-LHC CMS High Granularity Calorimeter (HGCal), $\ee$ collider alternatives (SiD and ILD for the ILC, 
the CLIC detector concept, CLD for the FCC-ee, and the baseline CepC detector concept), FCC-hh, and a possible muon collider. 
Among these, the CMS HGCal upgrade for the HL-LHC is at the most advanced stage of development, with delivery of the production 
silicon sensors slated to begin in January 2023. Another example of high granularity is the development of MAPS electromagnetic 
calorimetry for the SiD detector. This technique has shown the potential for excellent photon reconstruction and separation, and 
excellent $\pi_0$ reconstruction while delivering very good electromagnetic energy resolution. The addition of timing layers would 
also be a natural extension of the MAPS approach. 

PF calorimetry development has synergies with developments in other areas of HEP instrumentation. Silicon detectors, SiPMs, fast 
scintillators, and gas ionization detectors are all good candidate active media, depending on the use case. Smart, low-power, 
radiation-tolerant front end electronics are needed to realize compact designs. PF reconstruction provides a benchmark for optimizing 
novel computational methods, like the use of ML in triggering or particle reconstruction. The computational problem of 
PF reconstruction, in which charged and neutral hadrons, electrons, muons, hadronic tau decays, photons, and jets must be 
identified from associations between tracks and highly granular calorimeter deposits, may be particularly well suited to ML methods 
as opposed to a traditional deterministic algorithm. In turn, PF reconstruction requires performance advances that can speed up 
detector simulation or improve jet energy resolution. Breakthroughs in any of these related technologies can drive significant 
progress in PF calorimetry, be it cost reduction, simplification of engineering, increased radiation hardness, or improved 
physics performance.

To realize the full potential of PF calorimetry, and to ensure that it is ``shovel-ready'' when the next large experiment is approved, 
R\&D is needed to solve outstanding problems. Extensive beam testing, especially of large-scale technological prototypes, is needed 
to study integration issues and how they affect EM and hadronic energy resolution. As stated above, more research is 
needed in FE design, especially for high-occupancy and high-radiation applications like a future hadron or muon collider. 
Finally, the challenge of rethinking traditional assembly and quality control procedures to find solutions that scale to tens to 
hundreds of millions of channels cannot be overlooked. There is a growing need to incorporate lessons from product and process 
engineering into instrumentation labs, either through personnel decisions or by adapting the training of traditional physics 
graduate students and postdocs. Similarly, academic-industrial partnerships need to be formed with this scale in mind.

\subsection{Dual Readout}
\label{sec:calorimetry-DRO}

The DR approach to the precise measurement of jet energies makes use of the scintillation light and Cherenkov light from 
showers. The relative amounts of scintillation and Cherenkov light is used to correct the shower energy. The original approach to 
implementing DR calorimetry used a matrix of scintillating fibers and clear optical fibers (for the Cherenkov component) 
in a block of absorber material. However, with the reduced need for spatial associations (as in PF), DR can make 
beneficial use of an EM front section using homogeneous scintillating crystals which have excellent EM 
energy resolution. Development of DR systems has followed both of these approaches. In the second case, the separation of 
the scintillation and Cherenkov signals can be achieved by the use of optical filters and SiPM readout or exploitation of the 
different time structure of the two signals. Development and testing of large-scale systems is needed to demonstrate the feasibility 
of both the all-fiber and homogeneous EM section plus fibers approaches to DR.

Significant R\&D is still needed to fully realize the potential of DR. Large-scale prototypes need to be constructed to 
understand which parts of the engineering are cost drivers, which need further R\&D, and how a realistic system will perform. Such 
efforts have already begun in the context of $\ee$ circular collider studies. The Innovative Detector for an Electron-positron 
Accelerator (IDEA) fiber calorimeter considers a stacked capillary arrangement as well as a 3D-printed copper alveolar tower into 
which fibers are inserted. In the case of homogeneous scintillating crystals, a key open question is how to ensure that readout of 
the red end of the Cherenkov spectrum, far from the scintillation peak, will yield a large enough signal with which to perform the 
DR correction. Due to the large size of collider detectors, another important question is how to drive down the cost of 
high-performing crystal scintillators. In the long-term, research into new optical materials, such as photonic crystal fibers for 
increased Cherenkov light collection, quantum dot and semiconductor nanoparticle wavelength shifters, and fibers that maintain 
polarization information, can improve the performance or reduce the cost of DR fiber calorimeters.

The wide availability of high-performance SiPMs serves to simplify DR mechanical design and allows for channel granularity 
not initially envisioned. There is no reason that the advantages of PF and DR cannot be combined in certain 
applications to yield even more flexibility. Readout of each fiber by a SiPM opens the door to jet imaging and position reconstruction 
using algorithmic or deep learning techniques. This can enable gluon-jet discrimination, studies of jet substructure, or 
identification of tau leptons or boosted objects. Indeed, the extra FE information that DR relies on requires 
advances in electronics for efficient data processing and reduction. Areas of R\&D in this direction include system-on-chip waveform 
digitizers with real-time analysis, FPGAs on the FE, and digital SiPMs. With flexibly designed 
readouts, it may also be possible to use the time structure of the Cherenkov and scintillation signals to estimate the shower's EM 
fraction online and improve triggering.

\section{Readout Electronics and ASICs}

In pursuit of its physics goals, the HEP community develops increasingly complex detectors, with each subsequent
generation pushing to ever finer granularity. For example, the emergence and growing dominance of PF reconstruction 
methods has stressed the increasing importance of very fine detector granularity, not only for tracking detectors but also for 
calorimetry, and has also initiated the push to 4D-detectors that combine precision measurements of both spatial and time 
coordinates to their measurements of energy and momentum. While these developments have led to enormous growth in detector channel 
counts, the analog performance requirements for energy and momentum measurements, as well as spatial and timing precision, are being 
maintained or (more typically) even tightened.

These challenging detector requirements drive the corresponding development of readout electronics. A variety of factors, including 
the growing specialization of the functionality required, and the explosive growth in channel counts and typically modest (if any) 
increases in the material budget and the power and cooling budgets, lead to the increasing reliance on custom-developed ASICs. 
This trend is further exacerbated by challenges in the various experimental environments, including radiation hardness 
requirements, cryogenic and deep-cryogenic operations, and space-based detector systems. The development of custom ASICs in 
advanced technology nodes allows HEP detector subsystems to achieve higher channel density, enhanced performance, lower power 
consumption, lower mass, much greater radiation tolerance, and improved cryogenic temperature performance than is possible
with commercial integrated circuits (ICs) or discrete components. The higher level of integration also leads to fewer components and
fewer connections, leading to higher reliability, as required by experiments that can run for decades and that provide access to the 
on-detector electronics at most annually, and sometimes never.

This writeup summarizes the work of IF07, the Instrumentation Frontier's Topical Group for Readout Electronics and 
ASICs~\cite{https://doi.org/10.48550/arxiv.2209.15519}. 
The community efforts as part of IF07 were organized across seven white papers submitted to the Snowmass process. The 
first~\cite{whitepaper1} discusses issues related to the need to maintain the talented workforce required to successfully 
develop future HEP electronics systems, and to provide the appropriate training and recruitment. The remaining papers focused 
on electronics for particular detector subsystems or technologies, including calorimetry~\cite{whitepaper2}, detectors for 
fast timing~\cite{whitepaper4}, optical links~\cite{readout-links}, smart sensors using AI~\cite{whitepaper6}, and RF readout 
systems~\cite{whitepaper8}. In the following section, a brief overview is provided of each of the white papers in turn. A 
white paper on silicon and photo-detectors was originally considered. This content is now captured by other frontiers with 
contributions from ASICs and readout electronics.

There are some overarching goals for advancing the field of readout and ASICs. Efforts with broad impact include:
\begin{itemize}
 \setlength{\itemindent}{1.5em}
\item[{\bf IF07-1}]: Provide baseline support and specialized {\bf training opportunities} for the instrumentation work force to keep 
the US electronics instrumentation and ASIC workforce current.
\item[{\bf IF07-2}]: Improve mechanisms for {\bf shared access to advanced technology} providing broader access by the community to 
foster real exchange of information and accelerate development.
\item[{\bf IF07-3}]: Continue to develop methodologies to {\bf adapt the technology for operation in extreme environments}. Deep 
cryogenics, ultra-radiopure materials, radiation-harsh environments with limited power budget and long lifetime for all cases.
\item[{\bf IF07-4}]: Develop novel techniques to manage very high {\bf data rates}. Data reduction and optimization needs to be as 
close as possible to the generation point with acceptable power consumption.
\item[{\bf IF07-5}]: Create framework and platform for {\bf easy access to design tools}. Develop specialized online resources for 
the HEP community (e.g. system simulations and design repository). Provide the basis for true co-design R\&D efforts: from simulation 
to verification and implementation.
\end{itemize}

A brief overview of each of the IF07 white papers is provided below. A more detailed discussion of each of the various topics 
can be found in the white papers themselves, and in the references therein.

\subsection{Workforce and Training Needs}
 
Two decades ago as we embarked on the design of the LHC detector systems, \emph{on detector} readout was focused on the integration 
of custom sensors and multiple types of ASICs with communications rates at or below 100 Mbps. 
Significant effort went into the design of printed circuits on boards or flexible substrates. A new generation of ASICs were coming 
into use that could survive radiation tolerance levels consistent with the needs of LHC provided specialized layout techniques were 
employed. Designers were faced with learning new tools and rules for commercial ASICs that were attainable in a few months and system 
interfaces were being designed that allowed multiple institutions to design parts of a readout system nearly independently. As LHC 
systems were being placed into service commercial ASICs and communications systems were leap frogging forward and significant change 
was moving the state of the art forward faster than our communities and budgets could maintain pace with. Today's HL-LHC designs have 
been able to take advantage of advances that are still several technology generations behind the commercial state of the art to move 
what previously was multi-chip functionality onto a single silicon substrate to make systems on a chip that operate at much higher 
clock rates and can hold and selectively read out more data and contain most or all of the readout blocks for modules of thousands of 
small size sensors. The price for successful submission of these far more complex, low power integrated circuits has been the 
requirement for a workforce with knowledge of a broad set of new tools for design and verification as well as formalized design 
management and integration tools that do not allow new versions or blocks to compromise the overall performance or design progress of 
these now highly complex systems on a chip. This IC revolution has also resulted in complex new capabilities built 
into today's FPGAs that comprise a large part of the \emph {off-detector} DAQ systems designed over the past decade. 
Here too the required knowledge base has expanded sufficiently far so that coding is no longer an easy skill to learn for this or next 
generation designers. We foresee the need for continuing workforce training past the qualifying academic degrees to update commercial 
design skills. Many designs require HEP specialized knowledge to enable successful first time designs for extreme temperature, 
radiation and high-radiopurity environments. Our community needs to exploit the internet to provide an archive with searchable 
access to design examples from previous generations of detectors to replace the institutional knowledge passed down from previous 
teams. This is especially important given that the time between large detector system developments may exceed career lifetimes. In 
these extended interim periods it is recognized that core HEP instrumentation specialists need to update their skills with practical 
projects to ensure their familiarity with the evolving state of the art in ASIC designs. This can be encouraged by DOE-provided 
annual FOAs to support the design of service blocks that will be necessary in yet to be defined FE ASICs for next generation 
sensor arrays: High-speed communication links, data storage blocks, Phase Lock Loops (PLL), power converters and regulators etc. 
Silicon tested designs for next generation ASICs will both speed up design cycles and help maintain workforce skill levels 
consistent with current (at the time of need) ASIC design tool familiarity to minimize the number of design submissions required 
to assure reliable ASIC performance. We encourage the HEP academic community to provide instrumentation-based PhD degrees to 
formally make available expertise in the experimental HEP community. In addition it will be beneficial to have instrumentation 
conferences with training available to introduce new design approaches or technologies. This training should include continuing 
education certification to help qualify candidates for positions in experimental HEP. We see the recent support from DOE for 
instrumentation-based traineeship to be an important step towards having a better trained, better informed workforce not only for 
designing systems but also for the benefit of future peer-reviewed systems.
 
 \subsection{Calorimeter Readout Electronics}
 
Calorimeters will continue to serve key roles at future colliders, as well as in many other applications. The traditional challenges 
of calorimeter readout systems, including providing high precision energy measurements over a very wide dynamic range, are 
increasingly compounded by the demands of much finer granularity and correspondingly higher readout rates, as well as the demand of 
providing precision time measurements in the move toward 5D-calorimetry. The on-detector location of the FE electronics, 
necessitated by signal-to-noise and other requirements, imposes additional challenges, including tolerance to radiation or magnetic 
fields, reliability over long periods without maintenance, and power and cooling budgets.

Key innovations that are driving ongoing and future readout developments include PF algorithms, in which calorimeter 
measurements of energies are combined with the momentum measurements from tracking detectors, and DR detectors, which 
provide a more flexible combination of the EM and hadronic components of a shower. Both of these methods aim to 
significantly improve the energy resolution, and both can be implemented either with or without timing information as an added 
component. 

PFA are pushing calorimeter designs to ever finer granularities, and therefore greatly increasing channel 
counts. As an example, the HGCal being developed currently for the CMS HL-LHC upgrade
includes over six million readout channels, a dramatic increase over the $\approx 200k$ channels of the ATLAS LAr  
calorimeters that set the scale for "finely segmented" among the original LHC detectors. 

Meeting the challenges for FE calorimeter readouts has relied on custom ASICs for over 30 years, and ASICs will only become 
increasingly important. Fortunately, the higher level of integration available today has permitted some consolidation; for example, 
while the current ATLAS LAr FE required development of 11 different custom ASICs, spread over a variety of technologies, the HL-LHC 
developments underway requires only three. The higher level of integration is most clearly seen in the digital realm, where for 
example CERN's lpGBT chip in 65~nm CMOS fulfills a number of functions, including clock and control distribution, slow control 
monitoring, and data serialization and formatting, that were previously spread over a number of different ASICs. However, even in the 
analog realm some consolidation has been achieved, such as the 130~nm ASIC developed for LAr that combines the functions of both 
preamplifier and shaper. 

Preserving the ability to develop ASICs to meet the challenges of future calorimeters relies on maintaining affordable access to the 
specialized ASIC processes in industry, and to qualify these processes concerning radiation tolerance and, in some applications, also 
their performance in cryogenic environments. A fortunate development has been the typically increasing radiation tolerance of new ASIC
processes with smaller feature sizes; while the original LHC FE ASICs exploited a number of specialized processes, or used standard 
cells and design rules that had to be explicitly developed to improve the radiation tolerance, the HL-LHC developments can focus on
commercial 130~nm and 65~nm CMOS processes and use commercial standard cell libraries. The move to ever smaller feature-size ASIC 
processes also greatly helps reduce the power consumption. However, the corresponding evolution to lower power rails presents a 
significant challenge for the very FE analog circuits that must handle input signals over a very wide, often around 16-bit, dynamic 
range.

Other challenges for future calorimeter readouts include the powering systems, ranging from power supplies, to 
digital-current-to-digital-current (DC-DC) converters and low dropout regulators (LDO) or on-chip regulation, that can provide the 
needed power in a practical and radiation-tolerant manner. Powering has long proved a difficult issue at the LHC, and carefully 
evaluating over many years the radiation tolerance of commercial devices has clearly demonstrated that the great majority would 
not survive the LHC conditions. As a result, industrial partnerships were launched for both the original LHC and for the HL-LHC 
to develop radiation-tolerant LDOs. 

\subsection{Electronics for Fast Timing}
Picosecond-level timing will be an important component of the next generation of detectors. The ability to add an additional dimension 
to our measurements will help address the increasing complexity of events at hadron colliders and provide new tools for precise 
tracking and calorimetry for all experiments. Timing is crucial for background rejection in DM searches and neutrino 
detectors. As resolution continues to improve, time will likely become an equal partner to position measurement in tracking and 
PFA. All this has been enabled both by the rapid and continuous advance of fast electronics and the ability to 
generate fast signals with good signal-to-noise. Fast silicon sensors with gain, (e.g. LGAD) and sensors 
without gain (e.g. 3D-sensors), MPGD, Cherenkov light and fast scintillators such as Lutetium-yttrium oxyorthosilicate (LYSO) 
crystals, microchannel plate (e.g. LAPPD), semiconductor-based photodetectors such as SPADs and SiPMs, and other sensors all 
provide very fast signals. There are several R\&D efforts aimed at the development of fast ASIC electronics for future HEP 
applications with focus on using specific and advanced technology (e.g. silicon-germanium (SiGe), 28 and 22 nm CMOS) and 
implementing suitable concepts for the needed time resolution (e.g. full waveform digitization, TDC, monolithic solutions).

The design and optimization of the FE amplifier is particularly important for fast electronics. Distribution and maintenance of 
reference clocks will be a challenge for large systems. Timing must be monitored to compensate temperature and aging effects. 
Complex calibration techniques are being developed to provide 1~ps phase resolution. 

Using a combination of instrumentation techniques and developments, including Constant Fraction Discrimination (CFD), waveform 
sampling combined with precise clock distribution and newly developed sensors we expect that it will be reasonable for future fast 
timing detector sub-systems to set a 10 ps timing resolution goal that will allow for unprecedented accuracy and significantly 
improve the physics reach of next generation high rate, collider detectors. 

\subsection{Optical Links}
\label{sec:links}

The dramatic increase in channel count due to new fine-grained detectors relies on optical links to transmit data from the 
on-detector FE electronics to the TDAQ systems off detector. Today's high-radiation detector environments require custom ASICs and 
materials for these links. 

To date, these radiation requirements and reliance on custom solutions have resulted in link speeds per fiber that are significantly 
lower than used in commercial systems. Custom links developed for LHC were limited to less than 2~Gbps per fiber. Recently 
completed HL-LHC upgrades have pushed this to 10~Gbps. This substantial increase in per-fiber bandwidth has, however, not kept pace 
with the growth in data volume. The original ATLAS LAr readout used a single 1.6~Gbps link per 128 channels, while the corresponding 
HL-LHC readout will need 22 fibers at 10~Gbps each for 128 channels. These faster links are still well behind current technology that 
is pushing 56~Gbps. The 10~Gbps lpGBT development led by CERN in 65~nm CMOS will be used in most HL-LHC on-detector electronic
readout applications and R\&D is currently underway that aims to increase the bandwidth by a factor of two in the short term, perhaps 
in time for some HL-LHC sub-systems still within 65~nm CMOS. Studies are underway to push to 56~Gbps utilizing a 28~nm CMOS process. 
Realization of a functional optical link requires several building blocks, including a data fan-in and serializer, combined with a
electrical-to-optical conversion coupled to an optical module that couples to the fibers. Apart from the ASICs, the critical 
components for the optical modules themselves have been mostly satisfied by commercially available VCSELs and pin diodes, which
must be painstakingly selected for radiation tolerant materials. However, custom optical module designs that integrate these functions 
plus the optical connections are still required. These must satisfy tight spatial and mechanical constraints within location sensitive 
material budgets imposed, in particular, by the inner detectors at future high rate hadron colliders or at HL-LHC future upgrades. 

Meeting future needs will require ongoing R\&D, to facilitate the continued evolution to higher bandwidths, and in particular per-fiber 
bandwidths.

\subsection{Smart Sensors Using AI}

Modern particle physics experiments and accelerators create massive amounts of data which require real-time data reduction as close 
to the data source and sensors as possible. 

Data transmission is commonly much less efficient than data processing. Therefore, placing data compression, extracting waveform 
features and processing as close as possible to data creation while maintaining physics performance is a crucial task in modern 
physics experiments. The implementation of AI and ML in near-detector electronics is a 
natural path to add capability for detector readout. While the application of AI/ML is growing rapidly in science and industry, the 
needs of particle physics for speed, throughput, fidelity, interpretability, and reliability in extreme environments require advancing 
state-of-the-art technology in use-cases that go far beyond industrial and commercial applications.

Progress has been made towards generic real-time processing through inference including boosted decision trees and neural networks 
(NNs) using FPGAs in off-detector electronics, however ML methods are not commonly used to address 
the significant bottleneck in the transport of data from FE ASICs to back-end FPGAs. Embedding ML as close as possible to the data source 
has a number of potential benefits:

 1) ML algorithms can enable powerful and efficient non-linear data reduction or feature extraction techniques, beyond simple 
 summing and thresholding; 
 2) This could in turn reduce the complexity of down stream processing systems which would then have to aggregate less overall 
 information all the way to offline computing; 
 3) This enables real-time data filtering and triggering like at the LHC and the EIC 
 otherwise not be possible or be much less efficient; or in the case of cryogenic systems, reduce the system complexity;
 4) Furthermore, intelligent processing as close as possible to the source will enable faster feedback loops. 
 
Most of the efforts are focused on digital CMOS technology, such as implementations based on general-purpose Tensor Processing Units 
(TPU) or GPUs, FPGAs, and more specialized ML hardware accelerators. The steady improvements in such hardware platforms' performance 
and energy efficiency over the past decade are attributed to the use of very advanced, sub-10-nm CMOS processes and holistic 
optimization of circuits, architectures, and algorithms. The opportunities for building more efficient hardware may come from 
biological NN. Indeed, it is believed that the human brain, with its $>$1000$\times$ more synapses than the weights in the largest 
transformer network, is extremely energy efficient, which serves as a general motivation for developing neuromorphic hardware. There 
is a long history of CMOS neuromorphic circuits. However, unleashing the full potential of neuromorphic computing might require novel, 
{\bf beyond-CMOS} device and circuit technologies that allow for more efficient implementations of various functionalities of biological 
neural systems. 

Many emerging devices and circuit technologies are currently being explored for neuromorphic hardware implementations. Neuromorphic 
inference accelerators utilizing analog in-memory computing based on floating gate memories are perhaps the closest to widespread 
adoption, given the maturity of such technology, the practicality of its applications, and competitive performance with almost 1000x 
improvement in power as compared to conventional (digital CMOS) circuit implementations. The radiation hardness of these techniques 
and their applicability for robust performance in extreme environments is yet to be evaluated.

There are several ongoing R\&D efforts focused on on-detector AI/ML and the key elements of both the design and implementation, and 
the design tools themselves. Though early in this exploration, the existing work highlights the needs for configurable and adaptable 
designs and open-source and accessible design tools to implement efficient hardware AI. Different classes of future applications 
require dedicated investments to advance our capabilities in this field. 

\subsection{Cryogenic Readout}

Many applications, including direct DM detection, CMB and line-intensity mapping as well as neutrino experiments, require electronics designed to perform within cryogenic environments.

Some operate at deep cryogenic levels going from liquid helium (LHe) temperatures at a few Kelvin down to milli-Kelvin. TPCs used in 
neutrino detection and DM searches use noble liquids as a target material. Most current generation 
TPCs keep the active electronics in the warm, outside the cryostat. The sheer scale of the DUNE LAr TPCs being designed today makes 
it impractical to bring all the raw analog detector signals into the warm, and instead embeds the active FE electronics in 
the cold. For the DUNE wire-based LAr TPC readout, the required cold functionality is achieved by a set of three custom ASICs, one 
that amplifies and provides analog filtering, one that digitizes at 2~MSamples/s, and one that collects and serializes the data for 
transmission over copper cables to the warm electronics outside the cryostat.

The DUNE near detector will use a low-power pixel readout utilizing the self-triggering LarPix 64-channel pixel ASIC submerged in 
LAr. Four programmable readout paths on the chip allow a board-level fault tolerant readout path. Track reconstruction results from 
prototype boards populated with 100 ASICs each have shown very promising physics results. 
 
A second, very low-power readout is being investigated by the Qpix Collaboration that utilizes a novel approach to record the times 
of arrival of a fixed unit of charge, enabling offline reconstruction of the ionization current sensed at the pixel with a 
sub-femto Coulomb unit charge sensitivity.

Deep-cryogenic CMOS circuits and systems have seen rapid development in recent years, with quantum computing being the primary 
technological and economic driver. Early work focused on studying and modeling the properties of devices, including 
Metal-Oxide-Semiconductor Field-Effect Transistor (MOSFET), resistors, and capacitors, in various CMOS technologies at cryogenic 
temperatures. The results suggested that devices in modern nm-scale fabrication technologies (including bulk CMOS, 
SOI, and fin-shaped field-effect transistor (FinFET)) function successfully at temperatures as low as 50 mK. The main 
deleterious effects of cryogenic operation on nanoscale MOSFETs include a moderate increase in threshold voltage and 
degradation in 1/f noise and matching properties, both of which can be overcome using well-known precision circuit 
design techniques. A potentially more serious problem is the self-heating of such devices due to internal power dissipation. For 
example, a recent study shows that the channel temperature of 40~nm bulk CMOS transistors can increase by over 40~K compared to a 
LHe ambient of 4.2 K when dissipating only 2 mW. Thus, integrated electrothermal modeling is an important requirement for 
successfully designing deep-cryogenic CMOS systems. Given the availability of reliable cryogenic CMOS device models, research has 
focused on using these models to develop and test functional blocks necessary for qubit control and readout, including cryogenic 
frequency synthesizers, low-noise amplifiers, and circulators. Cryogenic operation of embedded memory, including standard static 
random access memory (SRAM) cells, has also been demonstrated. The availability of these building blocks has motivated the recent 
development of complete cryogenic CMOS qubit control and readout interface and monolithic quantum processors that integrate on-chip 
silicon charge qubits with CMOS readout electronics. Precision sensing is another major driver for cryogenic CMOS circuits and 
systems where ultra-low-noise analog FE is a major application area.

\subsection{RF Electronics}
RF electronics utilizes the RF domain to sense or control EM radiation well below ionization energies to make highly 
sensitive measurements of natural phenomena. The sophistication and sensitivity of RF-related detection systems has improved 
dramatically with advances in wireless communications. HEP now has the opportunity to probe EM field signals, measuring amplitude, phase, 
frequency, and polarization with unprecedented signal-to-noise and orders-of-magnitude improvement in signal processing per unit 
of power consumption. Combining RF analog-digital and digital-analog high-precision and bandwidth performance and high-speed FPGAs 
to read out arrayed superconducting sensors such as: TES, Kinetic Induction Detectors (KID) and superconducting bolometer systems will 
give us a better view of the universe. 
 
Measurements of the extreme red-shifted 21\,cm (1420 MHz) hyperfine transition of neutral atomic hydrogen are helping explore the 
early Universe in the red-shifted range of 10-500 MHz, inferring times before the creation of stars where a homogeneous mix of 
materials allow robust theoretical predictions to be tested. Sophisticated very low-noise techniques are required to achieve the 
necessary electronics noise levels. To achieve reasonable levels of detection, undistorted signal backgrounds can only be found 
locally on the far side of the moon. Multiple proposals are being developed with the objective of installing sensitive antennae and 
electronics there with lunar orbiting satellite relay stations. The results of these explorations will benefit the ongoing 
development of cosmological models and may be an indicator of new physics. 

\section{Noble Elements}
Particle detectors making use of noble elements in gaseous, liquid, or solid phases are prevalent in neutrino and DM experiments 
and are also used to a lesser extent in collider-based particle physics experiments. These experiments take advantage of both the very 
large, ultra-pure target volumes achievable and the multiple observable signal pathways possible in noble-element based particle detectors. 
As these experiments seek to increase their sensitivity, novel and improved technologies will be needed to enhance the precision of their 
measurements and to broaden the reach of their physics programs. The Priority Research Directions (PRDs) and thrusts identified in the BRN report~\cite{osti_1659761} are still relevant in the context of this Snowmass 2021 Topical Group. The areas of R\&D in noble element 
instrumentation that have been identified by the HEP community in the Snowmass white papers and Community Summer Study align well with the 
BRN report PRDs, and are highlighted by five key messages (with IF-wide themes in bold):
\begin{itemize}
 \setlength{\itemindent}{1.5em}
 \item[\bf IF08-1] Enhance and combine existing modalities (light and charge) to {\bf increase signal-to-noise and reconstruction fidelity}.
 \item[\bf IF08-2] {\bf Develop new modalities for signal detection} in noble elements, including methods based on ion drift, metastable 
 fluids, solid-phase detectors and dissolved targets. Collaborative and blue-sky R\&D should also be supported to enable advances in this area.
 \item[\bf IF08-3] Improve the understanding of {\bf detector microphysics} and calibrate {\bf detector response in new signal regimes}~\cite{https://doi.org/10.48550/arxiv.2203.07623}.
 \item[\bf IF08-4] {\bf Address challenges in scaling technologies}, including material purification, background mitigation, large-area readout, 
 and magnetization.
 \item[\bf IF08-5] {\bf Train the next generation of researchers}, using fast-turnaround instrumentation projects to provide the 
 design-through-result training no longer possible in very-large-scale experiments.
\end{itemize}

The IF08 Topical Group report~\cite{Dahl:2022bst} identifies and documents recent developments and future needs for noble element detector 
technologies. In addition, we highlight the opportunities that this area of research provides for continued training of the next generation of 
scientists. The following subsections summarize the objectives and opportunities associated with each of the above messages, described in detail 
in the Topical Group report.


\subsection{Enhancing Existing Modalities}
Noble element detectors developed for neutrino physics and DM searches record mainly the charge from the ionization 
electrons and the light from the scintillation produced by the passage of charged particles through the medium. The technologies to 
read out the charge in neutrino experiments have mostly been based on wire readouts, while charge measurements in DM  
searches rely on gain mechanisms such as gas electroluminescence and proportional gain. To read out scintillation light, both 
neutrino and DM experiments have focused on the use of Photo Multiplier Tubes (PMT) and SiPMs with coverage ranging from sub-percent to up to 
50$\%$ levels.

It is clear that for the future, advances in charge and light detection capabilities are highly desirable, and to this end a range 
of new approaches have been proposed. 

\subsubsection{Pixels}
Although the concept of wire-based readout has been proven and has had wide usage in neutrino detectors, it has an intrinsic 
limitation in resolving ambiguities, resulting in potential failures of event reconstruction. In addition, the construction and 
mounting of massive anode plane assemblies to host thousands of finely spaced wires poses significant and costly engineering 
challenges. For these reasons, a non-projective readout presents many advantages, but the large number of readout channels and the 
low-power consumption requirements have posed considerable challenges for applicability in liquid noble TPCs. The endeavour to build 
a low-power pixel-based charge readout for use in LArTPCs has independently inspired multiple consortia to pursue complimentary 
approaches to solving this problem.

\subsubsection{Light Collection}
The information carried by photons is critical for a wide range of physics measurements in noble element detectors, providing a 
crucial means by which to perform detector triggering as well as position and energy reconstruction and identifying interactions of 
interest. It is essential to fully leverage this information in next-generation measurements, including neutrino interactions from 
low-energy CEvNS up to the GeV energy scale, neutrino astrophysics, and Beyond the SM physics 
searches such as for low-mass DM and neutrinoless double beta decay. These efforts will require substantial (even as high 
as 100-fold) increases in light collection, to enable percent or sub-percent level energy resolution, mm-scale position resolution, 
low-energy detector readout triggering, or highly efficient PID, including for events around 
$\mathcal{O}(\leq 1)$~MeV (keV) energies in detectors at the 10~kton (100~ton) scale. In the context of the broader program of 
noble element detectors, enhancements in photon collection will lead to dramatic improvements in event reconstruction precision 
and PID, in a broader range of physics signatures afforded by lower trigger thresholds, and in precision timing 
to unlock new handles for beam-related events.

Measurement of the light signals, however, presents a major challenge with currently available technologies. For example, in 
large-scale liquid argon neutrino detectors, it is typical to collect $<$1$\%$ of the produced photons. This limitation is driven 
in large detectors by geometric considerations and other active components, total heat load, data volume, and the cost of 
instrumented surface area. Several promising approaches to improving light collection can address one or more of the limitations 
noted above, and taken together, provide a set of complementary tools for next-generation experiments: photon collection efficiency 
may be improved by imaging a large volume on a small active detector surface using novel lensing technologies, deployment of 
reflective and wavelength-shifting passive surfaces, bulk wavelength-shifting through dissolved dopants, improvement of photon 
detectors and photon transport efficiency, or the conversion of photons into ionization charge for readout using a TPC.

\subsubsection{Extreme Low Thresholds and Electron Counting}
The lowest energy phenomena that can be studied with existing noble-element modalities, including low-mass DM, reactor 
neutrinos, and natural (e.g. solar) neutrinos, require detectors that are sensitive to single ionization electrons. Such detectors 
are sensitive to $\mathcal{O}$(10~eV) electronic recoils and $\mathcal{O}$(100~eV) nuclear recoils, but lack the 
scintillation-dependent nuclear and electronic recoil discrimination present at higher energies. Without nuclear and electronic 
recoil discrimination, systematic backgrounds and radioactivity obscure typically background-free nuclear recoil event searches, 
and the radiopurity of detector materials becomes critical. Beyond background particle interactions, high rates of single- and 
few-electron signals are observed. Such spurious electrons have defied clear explanation and appear related to charge build-up on 
surfaces or in unknown chemical interactions, among other potential effects. Dedicated R\&D is needed to better understand the 
sources of these backgrounds and to develop mitigation techniques.

\subsubsection{Charge Gain}
Lower detection thresholds in track-reconstructing detectors (where the electroluminescence techniques of the previous section are 
less useful) may be achievable through amplification of the ionization signal in the form of charge gain, typically achieved in the 
gaseous phase of argon and xenon detectors. Multiple innovative methods are being developed to either enhance the capabilities of 
charge amplification in gaseous detectors (e.g. to achieve stable charge gain while retaining the primary scintillation channel) or 
to enable amplification directly in the liquid phase.

\subsection{New Modalities}
As a detection medium, noble elements present unique opportunities beyond the collection of scintillation photons and ionized 
electrons. The modalities described in this section find new ways to utilize the monolithic, ultra-pure elemental detection medium 
provided by noble elements, extending the reach of noble-element-based detectors to new signal regimes and enabling new methods of 
background discrimination in rare event searches.

\subsubsection{Ion Detection and Micron-scale Track Reconstruction}
The ability to reconstruct ionization tracks at micron- and sub-micron spatial resolution is the key to many currently unsolved 
detector challenges, including directional DM detection ($\mathcal{O}$($10^{-6}$g/cm$^2$) spatial resolution required), 
discrimination of single-electron backgrounds in 0$\nu\beta\beta$ searches ($\mathcal{O}$($10^{-1}$g/cm$^{2}$) spatial resolution 
required), and potentially for detection of supernova and solar neutrino events in very large-scale neutrino detectors 
($\mathcal{O}$(sub-mm) spatial resolution required). Attempts at direct (TPC-style) high-resolution reconstruction universally rely 
on ion drift rather than electron drift to escape the resolution-limiting effects of electron diffusion over large drift distance.

\subsubsection{Metastable Fluids}
Metastable fluid detectors amplify the energy deposited in particle interactions with the stored free energy in a superheated or 
supercooled liquid target. That amplification can be made selective by matching the different energy-loss mechanisms and length 
scales for signal and background interactions with the relevant phase-change thermodynamics, for example allowing bubble chambers to 
detect $\mathcal{O}$(1 keV) nuclear recoils (e.g. from DM or coherent neutrino scattering) while being completely blind to 
electron recoil backgrounds. Instrumentation efforts in this area typically focus on (i) extending phase-change based discrimination 
to new signal regimes, and (ii) improving control of spurious phase-change nucleation to enable larger quasi-background-free exposures.

\subsubsection{New Modalities in Existing Noble Element Detectors}
It is highly desirable to find novel ways to take advantage of both the field-wide expertise that has developed around 
noble-element-based detectors and the world-class infrastructure surrounding existing and near-future searches for new physics. In 
general, after an experiment has achieved its scientific goals, the experimental community can continue to leverage the infrastructure 
for future experiments. There are many liquid noble installations around the world that can lend themselves to this sort of upgrade. 
Specific past examples and future opportunities appear in~\cite{Dahl:2022bst}, including the use of dissolved targets in existing 
detectors and the development of a solid-phase TPC. 

\subsection{Challenges in Scaling Technologies}
Next-generation large-scale detectors are planned to search for DM and 0$\nu\beta\beta$ and to study neutrinos from both 
artificial and natural sources. Achieving these goals generally requires (i) scaled-up target procurement and radiopurity and purification 
capabilities; (ii) large area photosensor development with low-noise and low-radioactivity; (iii) high voltage and 
electric field capabilities compatible with multi-meter drifts and low-emission, large-area electrodes; and (iv) studying the 
effects and techniques for operating large doped noble liquid or gas detectors. The discovery capabilities of these detectors could 
be extended further by coupling them with a magnetic field, such to enable charge discrimination and improve momentum measurement.

\subsection{Cross-cutting Challenges}
In order to be sensitive to a wide range of physics phenomena, we must be able to make accurate and precise measurements of charge, light, 
or heat, from interactions of interest within our detectors, which requires in turn, a good understanding of the inherent noise levels, 
calibrations, and microphysics associated with these gaseous and liquid noble detectors. The challenges therein are cross-cutting, touching 
many different areas of experimental physics. But this also means that we can take a wider view to leverage facilities that will benefit 
many experiments across multiple frontiers. This deep understanding of the detection characteristics and calibrations will be essential 
components of preparing next-generation gaseous and liquid noble detectors for cutting-edge physics measurements in the Neutrino Physics 
Frontier, the Cosmic Frontier and the Rare Processes and Precision Measurements Frontier.

Flexible user facilities (not tied to any particular group nor experiment) play a key role in noble element R\&D, both for 
calibration needs where in-situ measurements are insufficient and for short-term instrumentation tests where a permanent dedicated 
test stand is unnecessary. These facilities minimize duplication of efforts by providing community-wide resources to benefit 
multiple research efforts. Laboratory-scale facilities such as the Liquid Noble Test Facilities at FNAL and Stanford Linear 
Accelerator Laboratory (SLAC) serve a broad, unique and important role in service of the liquid noble detector development 
community, both lowering the cost of entry for developing new techniques and testing prototypes, and providing an excellent 
training ground for students and postdocs. In an era of increasingly large-scale experimental programs, these facilities and 
the efforts they support provide much needed opportunities for junior personnel to design and build whole experiments, while 
gaining valuable technical expertise from collaboration with the facilities' engineers and technical staff.

\section{Radio Detection}
Detection techniques at radio wavelengths play an important role in the future of astrophysics experiments. The radio detection of cosmic 
rays, neutrinos, and photons has emerged as the technology of choice at the highest energies. Cosmological surveys require the detection 
of radiation at mm wavelengths at thresholds down to the fundamental noise limit. A summary of the Snowmass Instrumentation Frontier's 
Topical Group on Radio Detection can be found in~\cite{https://doi.org/10.48550/arxiv.2209.00590}.

High energy astro-particle and neutrino detectors use large volumes of a naturally occurring suitable dielectric: the Earth's 
atmosphere and large volumes of cold ice as available in polar regions. The detection technology for radio detection of cosmic 
particles has matured in the past decade and is ready to move beyond prototyping or mid-scale applications. Instrumentation for 
radio detection has reached a maturity for science scale detectors. Radio detection provides competitive results in terms of the 
measurement of energy and direction and in particle identification when to compared to currently applied technologies for high-energy
neutrinos when deployed in ice and for ultra-high-energy (UHE) cosmic rays, neutrinos, and photons when deployed in the atmosphere. It 
has significant advantages in terms of cost per detection station and ease of deployment.

\begin{itemize}
 \setlength{\itemindent}{1.5em}
\item [\bf IF10-1] While conceptual designs exist for next-generation arrays, investment in R\&D can reduce cost and optimize 
designs.
\item [\bf IF10-2] Opportunities exist in optimizing for power and simplifying: e.g. by using ASIC-based digitizer and readout or 
RF System on Chip (RFSoC). Investigations are needed to identify synergies with experimental needs in other areas.
\item [\bf IF10-3] Remote power and communications approaches of very large extended arrays can still benefit from dedicated R\&D. 
Explore synergies with experiments in other areas, like with the DUNE, SKA, and CTA experiments, which also need large distance 
communication and synchronization.
\item [\bf IF10-4] Enable future mm-wave cosmic probes through R\&D and pathfinder experiments of new mm-wave detectors with higher 
channel density relative to current Cosmic Microwave Background (CMB) detectors.
\end{itemize}

\subsection{Radio Detection of Cosmic Particles}

Astrophysical neutrinos, cosmic rays, and gamma rays are excellent probes of astroparticle physics and HEP~\cite{NF-04-Ackerman}. 
High-energy and UHE cosmic particles probe fundamental physics from the TeV-scale to the EeV-scale and beyond.

Radio detection offers opportunities to instrument large areas on the surface of the earth. This is important because at energies 
above an EeV, the cross section for neutrinos interacting with matter have increased so much that the detector area becomes a more
important parameter for the acceptance than the fiducial volume. At these energies, the preferred conversion targets for the neutrinos
are mountains, the earth's crust, ice, or even the atmosphere. For UHE cosmic rays and photons, radio detection 
utilizes the air shower formed in the atmosphere.

While the optical Cherenkov technique has been enormously successful in measuring neutrinos from $\mathcal{O}$({MeV}) to 
$\mathcal{O}$(10\,{PeV}), at higher energies, techniques at RF are the most promising for neutrino detection. In the 
energy range from 10 TeV to 30 PeV a factor of 5 in increase of sensitivity is needed in order to move from the discovery of cosmic 
neutrinos and the first correlation with sources to identify the sources of the cosmic neutrino flux and explore fundamental physics 
with high energy neutrinos. 

Like in the optical, for radio techniques the two media being pursued broadly are the atmosphere and solid natural ice, the latter 
with emphasis on neutrinos. Both can be used to pursue science with cosmic rays, energetic neutrinos, or both. Above about 100\,PeV, 
volumes of order 1000\,km$^{3}$, which can only be naturally occurring, are necessary to detect the rare astrophysical neutrino flux 
in this regime. Volumes of this scale are too large to instrument with the tens-of-km spacing necessary for optical sensors. This 
spacing of optical is set by the distance over which optical light is absorbed or scattered in the glacial ice that is used as the 
detection medium. However, clear ice is transparent to RF signals over distances of order 1\,km, which sets the 
typical spacing for detectors using radio techniques.

There are many different approaches to the detection of cosmic particles at RF. These approaches, the instrumentation 
that is unique and common to each, and future developments are described in this section.

\subsection{Askaryan}

In matter, particle cascades induced by cosmic particles produce Askaryan emission. The emission comes about from the time evolution 
of a charge asymmetry that develops in these particle cascades by ionization electrons moving with the cascade's front and positive 
ions staying behind. It is coherent for frequencies up to about 1\,GHz when viewed at the Cherenkov angle and is linearly polarized 
perpendicular to the cascade direction and inward to its axis. Askaryan emission was measured in test beam experiments in the 2000s. 
It has also been observed to be emitted from cosmic ray air showers. Now, many experiments are searching for this signature from 
UHE neutrino interactions either from within the ice or by viewing ice from an altitude.

\subsubsection{In-ice}

One approach to detecting neutrinos via the Askaryan signature is by embedding antennas into the ice itself, and the first 
experiment to take this approach was RICE. Since then, ARIANNA, ARA, and RNO-G search for UHE neutrinos with antennas 
deployed in the ice, either at shallow depths (within $\sim$10\,m of the surface) or somewhat deeper (up to 200\,m). An integrated 
facility for a wide band neutrino detector is IceCube-Gen2 that will employ both optical and radio detectors. Deep detectors are
employing interferometry-based trigger designs.

\subsubsection{Balloon}

Another approach to detecting the Askaryan signature from UHE neutrinos is to deploy antennas on a balloon-borne
payload at stratospheric altitudes where the payload can view approximately 1.5\,million km$^2$ of ice. At these altitudes, the 
threshold for a detection is higher, but above threshold, balloons greatly exceed ground-based approached in viewable area. The 
ANITA project flew four times under the long-duration balloon program of the National Aeronautics and Space Association (NASA), 
and the next-generation PUEO is set to launch in the 2024-2025 Austral summer season with a trigger that will use an interferometric 
approach.

\subsection{Geomagnetic Emission}

A cosmic ray incident on the atmosphere will produce a particle cascade, and charges in the cascade produce a transverse current 
due to the earth's magnetic field, which produces what is known as geomagnetic emission at RF. The geomagnetic 
emission is coherent up to frequencies of about 1\,GHz and is linearly polarized perpendicular to both the earth's magnetic field 
and cascade direction. The Askaryan emission mentioned above also contributes to the RF signal from air showers on 
average at the 20-30\,\% level in energy density. The interplay between geomagnetic and Askaryan emission gives rise to a 
non-azimuthally symmetric signal pattern due to the interplay of the different radiation patterns. Both are collimated in a Cherenkov 
cone with an opening angle of typically a degree.

A next generation UHE cosmic particle observatory is required to exploit proton astronomy and to discover 
UHE neutrinos and photons. Such a next-generation cosmic particle observatory needs to cover a large area, up to 
200,000\,km$^2$. Radio detection is the current favorite to make deployment at this scale possible and the GRAND experiment was 
proposed to cover an area of this size.

\subsubsection{Radio Detection of Cosmic Rays}

In addition to the traditional strategies for detecting high energy cosmic rays incident on the atmosphere, which include detection
of secondary particles, optical Cherenkov emission, and fluorescence emission, geomagnetic emission is a strategy that complements 
the others and has seen major advancements in the past two decades. Radio detection of cosmic rays has a round-the-clock duty cycle 
and leaves a broad ($\sim 100$\,m) footprint on the ground.

\subsubsection{Detection of Neutrino-induced, Earth-skimming Air Showers}

The combined geomagnetic and Askaryan signatures can also be used to detect air showers that can originate from energetic tau 
neutrinos that are earth skimming~\cite{CF-07-Abraham}. These can produce a tau lepton through a charged current interaction in 
matter, and the tau can subsequently decay to produce an EM or hadronic shower.

The detection of pulses consistent with air showers going in the upward direction has the advantage of being flavor sensitive, and 
has become an objective of many detectors. BEACON, currently in the prototype phase, TAROGE, and TAROGE-M are compact antenna arrays 
in elevated locations that aim to detect UHE $\nu_{\tau}$ emerging upwards via the radio emission of the air showers that they trigger. ANITA 
and PUEO are also sensitive to upgoing $\nu_{\tau}$, from a higher elevation. GRAND is a planned experiment that will cover large 
areas with a sparse antenna array to detect the radio emission from air showers triggered by UHE $\nu_{\tau}$, cosmic rays, and 
gamma rays. This is also a neutrino signature for PUEO, the follow-up of ANITA, and many other dedicated projects instrumenting 
large areas with antennas such as GRAND or putting them in mountains such as BEACON and TAROGE.

\subsection{Radio Detection and Ranging (RADAR)}

An alternate strategy for detection of neutrinos at RF uses RADAR. When a high-energy neutrino interacts in the ice, 
it produces a relativistic cascade of charged particles that traverse the medium. As they progress, they ionize the medium, leaving 
behind a cloud of stationary charge. This cloud of charge, which persists for a few to tens of nanoseconds, is dense enough to 
reflect radio waves. Therefore, to detect a neutrino, a transmitter can illuminate a volume of dense material like ice, and if a 
neutrino interacts within this volume, the transmitted radio will be reflected from the ionization cloud to a distant receiver, 
which monitors the same illuminated volume.

With this technique, a custom signal is transmitted in the ice and received after reflections from neutrino-induced cascades. The 
experimenter can thus determine the properties of the signal (including the amplitude, up to what is permitted to be 
transmitted). Also, the RADAR method has excellent geometric acceptance relative to passive (Askaryan) methods, which require the 
detector to lie within a small angular window at the Cherenkov angle. Recent test beam measurements have demonstrated the feasibility 
of the method in the laboratory, with in-situ tests forthcoming. RET-CR will serve as a pathfinder experiment, and RET-N could enable  
radio detection of UHE neutrinos within the decade with the potential to complement or improve upon existing technologies in this
energy regime.

\subsection{KIDs}

Cosmological surveys require the detection of radiation at mm wavelengths at thresholds down to the fundamental noise limit. The 
detection of mm-wave radiation is important for: studying cosmic acceleration (Dark Energy) and testing for deviations from general 
relativity expectations through measurements of the kinetic Sunyaev-Zeldovich effect, precision cosmology (sub-arcminute scales) 
and probing new physics through ultra-deep measurements of small-scale CMB anisotropy, and mm-wave spectroscopy to map out the 
distribution of cosmological structure at the largest scales and highest redshifts.

Imaging and polarimetry surveys at sub-arcminute scales will require O($10^6$) detectors over a O(10 deg$^2$) fields-of-view (FoV) 
covering nine spectral bands from 30\,GHz to 420\,GHz. Spectroscopic surveys (over a smaller FoV initially, O(1\,deg$^2$), but 
potentially also reaching O(10\,deg$^2$)) will require a further factor of 10–100 increase in detector count. The 2019 report of the 
DOE Basic Research Needs Study on HEP Detector Research and Development identified the need to carry out detector 
R\&D to achieve this goal. The driver for new technology is not the detector count but the increased detector {\it density}. Whereas 
in current experiments, the detector packing density is limited by the physical size of elements in existing demonstrated 
multiplexing schemes, KIDs eliminate the need for additional cold multiplexing components, allowing for arrays at the densities 
needed for the science aims of proposed cosmological surveys.

KIDs are a technology that has gained significant traction in a wide range of applications across 
experimental astronomy over the last decade. A KID is a pair-breaking detector based on a superconducting thin-film microwave 
resonator, where the relative population of paired (Cooper pairs) and un-paired (quasiparticles) charge carriers govern the total 
complex conductivity of the superconductor. Photons with energy greater than the Cooper pair binding energy (2$\Delta$) are able to 
create quasiparticle excitations and modify the conductivity. By lithographically patterning the film into a microwave resonator, 
this modification is sensed by monitoring the resonant frequency and quality factor of the resonator. Since each detector is formed 
from a microwave resonator with a unique resonant frequency, a large number of detectors can be read out without the need for 
additional cryogenic multiplexing components. In addition, the designs to be fabricated are relatively simple.

There are a number of KID-based architectures being developed for a variety of scientific applications. Direct Absorbing KIDs are
the simplest variant, where the resonator geometry is optimized to act as an impedance-matched absorber to efficiently collect the 
incoming signal. To date, the only facility-grade KID-based instruments are based on this detector architecture, with the NIKA-2 
experiment on the IRAM 30-meter telescope having demonstrated that the KID-based instruments are highly competitive with other 
approaches. Microstrip-coupled KIDs take advantage of recent advancements that allow for the ability to lithographically define 
circuits capable of on-chip signal processing with extremely low loss. The capability to robustly couple radiation from 
superconducting thin-film microstrip transmission lines into a KID with high optical efficiency is being developed. Thermal KIDs 
take a similar approach as has been developed for bolometric TES arrays. Instead of directly 
absorbed radiation breaking pairs, a thermally-mediated KID (TKID) uses the intrinsic temperature response of the superconducting 
film to monitor the temperature, and therefore absorbed power. It combines the multiplexing advantage of KIDs with the proven 
performance of bolometric designs in TES detectors, at the expense of fabrication complexity.

On-chip spectroscopy is a natural extension of multi-band imaging using on-chip filters to a filter-bank architecture to realize 
medium-resolution spectroscopic capability. Several approaches to on-chip spectroscopy exist at a range of technological readiness.

\section{Cross-cutting Themes in Support of Detector R\&D}
The IF09 Topical Group has identified several cross-cutting themes that are crucial to address, as they affect almost all aspects of 
carrying out successful Detector R\&D mentioned in the technical Topical Groups. They mostly deal with common infrastructure, such 
as detector facilities, access and to industry, workforce development and funding and collaborative models. The key findings are 
summarized here, while additional details are given further down.

\begin{itemize}
 \setlength{\itemindent}{1.5em}
\item [\bf IF09-1] Presently US funding for advanced detector R\&D is institute-based rather than collaboration-based. Yet 
collaborations are more essential than ever to leadership in detector R\&D technology. To a significant extent, funding constraints 
have limited the opportunity to establish significant collaborative detector R\&D programs. We recommend that funding for the DOE's 
KA25 Detector R\&D program should be increased significantly (2x-5x) to enable the establishment of collaborative R\&D programs that 
can address Grand Challenges and pursue blue-sky concepts. 

\item [\bf IF09-2] Both, retention of key experts (institutional memory) and workforce development are critical to detector R\&D. 
The recent “DOE traineeship in HEP instrumentation” is a step in the right direction and should continue. However, this 
focuses on the physicists while the instrumentation enterprise relies on a large set of uniquely skilled supporting players -- 
mechanical and electrical engineers, composites designers and fabricators, device physicists, chemists, materials scientists and 
technicians. Development of the next generation of all of these experts is vital. We recommend that funding for key technical 
personnel essential to these capabilities be considered on equal footing with the physicists.

\item [\bf IF09-3] Innovative instrumentation research is one of the defining characteristics of the field of particle physics. 
Blue-sky developments in particle physics have often been of broader application and had immense societal benefit. It is essential 
that adequate resources be provided to support more speculative blue-sky R\&D. One difficulty in the existing funding system is 
comparative review of incremental and blue-sky proposals. A separate funding opportunity and review process for blue-sky proposals 
could lead to more innovative R\&D activity.

\item [\bf IF09-4] New R\&D frameworks could enhance development: examples are the very successful CERN RD collaborations and the 
National Nuclear Security Administration's (NNSA) NA-20 consortia. LAPPD development was a HEP example that worked like that, but 
that has not been the detector R\&D typical model. The ECFA detector roadmap recommended establishing six new RD Collaborations. 
The US should engage broadly and deeply to shape these global instrumentation RD collaborations. We recommend the US HEP community 
establish a robust collaborative research program that includes both participation in the international RD collaborations and 
establishing domestic detector R\&D consortia.

\item [\bf IF09-5] Facilities (mainly at National Labs) are a vital element of detector technology development and should be supported. 
Present gaps that could be strengthened include: high-quality electron test beams, multi-TeV test beams, user access to 
low-temperature facilities covering liquid noble elements down to mK temperatures, low-noise - including vibration, RF, radioactive and 
cosmics - high-dose irradiation, and foundry access for radiation-hard microelectronics, semiconductor detectors, and superconducting 
devices for both development and production.

\item [\bf IF09-6] Multidisciplinary problems are prevalent in many areas of instrumentation development. The topic of 
multidisciplinary R\&D was explored in the Snowmass MultiHEP 2020 workshop. Multidisciplinary work has funding, recruiting, and 
career development challenges that new R\&D frameworks could alleviate.

\item [\bf IF09-7] Several themes that cut across multiple Instrumentation Frontier working groups were identified. The community 
should find ways to share expertise, tools, and developments in these areas: fast timing, cryogenic operation, and the need to solve 
challenging materials science and chemistry problems. 
\end{itemize}

\subsection{Facilities}
The HEP detector instrumentation community depends on facilities and capabilities that are beyond the scale of the individual 
researchers' grasp. Those include test beam, irradiation facilities, cryogenic facilities, semiconductor foundry access, 
low-background and underground facilities, etc. 

\subsubsection{Calibration and Test Beam Facilities}
Progress in particle physics depends on a multitude of unique facilities and capabilities that enable the advancement of detector 
technologies. Among these are test beams and irradiation facilities, which allow users to test the performance and lifetime of their 
detectors under realistic conditions. Test beam facilities are particularly important for collider and neutrino detector applications, 
while irradiation facilities are crucial for collider as well as some space-based astro particle detectors. It is important that the 
energy, intensity, particle composition and time structure of the beams are adequate for the detector needs for the next generation. 
In the area of irradiation facilities overlapping needs between detector and accelerator instrumentation as well as targetry 
development can be addressed.

To establish the global status of calibration and test beam facilities, the {\it Snowmass IF09 Test and Calibration Beams and 
Irradiation Facilities workshop} was held~\cite{beam}. A summary of the workshop can be found in a Snowmass white paper~\cite{Hartz} 
where we also develop the need and proposals for future facilities.

As summarized in the recent report on Basic Research Needs for HEP Detector Research and Development~\cite{osti_1659761}, 
future particle accelerator-based experiments will experience unprecedented radiation exposures of 
10 GigaGray ionizing dose, and fluences of $ {\rm 10^{18}~ 1~ MeV~ neq/cm^2}$ over their lifetimes. Detectors, support structures, 
electronics, data transmission components, and on-board data processing units will all need to be evaluated for performance at these dose 
rates and integrated doses two orders of magnitude greater than systems operating today. This will require new high-dose-rate environments 
for accelerated testing so that the anticipated integrated dose can be delivered in days rather than months or years. Activation of the 
detector materials will limit facility throughput unless remote material handling techniques are implemented. Such techniques are already 
employed for irradiation and post irradiation evaluation of accelerator components and for targets and windows for neutrino beam lines.

A broad array of test beam capabilities are also required to develop the next generation of detectors for tracking, calorimetry, and 
particle identification. In particular, access to beams of different particle species at a wide range of energies is critical to test the 
response and performance of different detector technologies, where well-calibrated instrumentation such as trigger hodoscopes, tracking 
telescopes, and Cherenkov detectors is often required to analyze and understand the performance of the device under test. Meanwhile, some 
of the capabilities being pursued for next-generation detectors place stringent demands on test beam time structure. For example, testing 
detector performance at high event rate and extreme (ps-scale) time resolution require test beams with high repetition rate and extremely 
short pulses, respectively. The BRN report calls for “a significant revitalization of US [test beam] facilities...to 
enable the detailed tests required” and highlights very precise test beams (in multiple dimensions -- spatial, energy spread, timing, and 
intensity) as an important area where US-based facilities can play a world-leading role.

Meeting the needs of this diverse community motivates the support of a variety of test beams, with different features and 
capabilities, across the DOE complex as well as the use of international facilities. The white paper~\cite{Hartz} surveys test-beam needs 
within the HEP instrumentation community and describes several opportunities for new facilities, improvements to existing test-beams, and 
detector test platforms that will provide essential and mutually complementary support for instrumentation development. Prospects include: 
plans and applications for a high-intensity proton irradiation facility at FNAL; a new facility with improved infrastructure for test 
beam experiments at FNAL; a new multi-purpose electron beamline at SLAC capable of delivering high repetition rates and short 
pulses with precise, flexible timing for test beam applications; and the Water Cherenkov Test Experiment (WCTE) at CERN. Further into the 
future in anticipation of new accelerators at the high energy frontier multi-TeV test beams will be necessary.

\subsubsection{Low-noise and Environmentally Stable Facilities}
The particle physics community relies on a variety of specialized environmental conditions for operation of the most sensitive instrumentation. 
These include low-vibration and geologically quiet environments, low EM interference environments, low-radioactive background 
environments, and cryogenic environments to suppress thermal noise. Environmentally stable facilities, i.e. low vibration or low 
EM interference, tend to be very specialized for specific equipment (e.g. isolating a laser system or Nuclear Magnetic Resonance 
(NMR)) or to suppress specific frequency bands (seismic or EM) making it difficult to define the requirements for a general 
purpose user facility. In addition, these facilities are generally within the reach of individual research groups to establish 
locally as needed.

\subsubsection{Cryogenic Test Facilities}
Cryogenic test facilities are critical infrastructure for physics experiments in a variety of fields. Some notable examples include studying 
noble liquid properties for particle detection, low-temperature device development, and research in quantum information. The required technical 
knowledge and infrastructure capacity, including the cost of setting up and operating the test facilities, can place them out of reach of many 
individual research groups. As such, these research areas would greatly benefit from centralized user facilities. Cryogenic test facilities 
can require significant investments in infrastructure and ongoing operations costs, and warrant consideration of the development of national user 
facilities. 

Several international institutions have established LAr test stands typically for specific purposes or experiments with limited measurement 
capabilities and access to the community. Some relatively large-sized facilities (more than 100 liters of LAr) include the ICARUS $\sim 50$ 
liter LArTPC at CERN~\cite{I9-Arneodo} with $\sim250$ liter LAr capacity, which is mainly used for ICARUS and DUNE detector electronics readout 
studies; the ArgonCube detector at University of Bern with $\sim1100$ liter LAr capacity, which is dedicated for pixel readout in LAr and DUNE’s 
modular detector study; and the Integrated Cryostat and Electronics Built for Experimental Research Goals (ICEBERG) at FNAL with $\sim3000$ 
liter LAr capability, which is dedicated to DUNE cold electronics studies. As can be seen, these facilities are typically tied to the specific 
tasks that are not easily accessible as a user facility with the general purpose of R\&D, such as testing new devices and detector designs, 
studying noble liquid property, calibrating detector signal response, or studying light detector performance, etc. The necessary initial 
components for a community LAr test facility include a large cryostat with sufficient volume and drift distance, adequate cryogenic 
infrastructure for condensing and gas circulation, a purification system to remove impurities with high electron attachment, a purity 
monitoring system with gas analyzers, basic high voltage system and readout electronics for TPC measurements, a photon detection system 
to study scintillation light, and the accompanying DAQ system.

The operation of sensors and systems at ultra-low temperatures is a major growth field, but one which necessarily has a high barrier for entry due 
to the relatively high cost of equipment, not just for the cooling platform itself but also associated measurement equipment and electronics and 
a knowledge gap in the operation and engineering of devices and systems at cryogenic temperatures. In addition, the typical fabrication, 
measurement and analysis cycle associated with the development of cryogenic devices will often mean that test stands can be idle for 
considerable periods of time, making the economics of every PI or research group having dedicated research facilities economically unattractive. 
The establishment of a centralized facility would democratize the process of device development by allowing users access to test stands and 
measurement equipment for a variety of different tests, including but not limited to thermal, RF and low frequency tests. This is particularly 
valuable at the pre-proposal and proof-of-concept stages to validate new ideas before committing to full proposals or the expense of purchasing 
dedicated test equipment. In addition, such a facility would serve as a repository of knowledge of low temperature materials and other specialized 
information and offer a medium to connect researchers with specialized cryogenic engineering resources. Finally, such a facility is a vital tool 
for training a future workforce by providing hands-on experience in the operation of cryogenic systems and measurement equipment, typically 
outside the ability of a University level teaching laboratory to provide. Such a facility would necessarily consist of more than a single 
refrigerator test stand, since this is something that a well-equipped PI laboratory would be able to provide. Instead, it is envisaged that the 
proposed user facility would provide a suite of test stands available for different experiment types, including a number of test stands providing 
environments for the testing of specialized devices such as in underground and low-background environments, test stands with optical access, and 
stands incorporating magnetic fields. An initial set of capabilities for the facility would include DC and RF measurements with optional optical 
access and in the presence of a magnetic field or in a low-background (underground) environment. It should provide fast turnaround with minimal 
DC and RF screening capabilities as well as a platform for training of operators.

One facility that should be mentioned in this context is the Cryogenic Underground Test Facility (CUTE) at SNOLAB~\cite{CUTE}. CUTE has already 
been used to validate performance of SNSPDs in a low-background setting, and many of its features seem to be addressing those needs described 
above. We would like to suggest the US HEP community to find ways to engage in this facility to the extent possible. 

\subsubsection{Semiconductor and Superconductor Foundry Access}
HEP experiments rely on foundry access for production of custom sensors and electronics. Semiconducting sensors include silicon sensors 
(strips and pixels, MAPS etc.), CCD and CMOS imagers, and silicon and germanium bolometers. Superconducting sensors (TES, MKID, SNSPDs, etc.) 
are used as particle detectors, microwave sensors, and optical imaging arrays. Specialized HEP electronics typically take the form of ASICs, 
often operating in atypical environments including high radiation dose areas and at cryogenic temperatures, but also include devices 
incorporating superconducting materials (SQUIDS, qubits), and III-V semiconductors (e.g. high-electron-mobility transistor (HEMT) amplifiers). 
Looking further ahead, new processor technologies for on-board trigger and data acquisition for high-rate experiments may require 
purpose-built microelectronics. Access to foundry services has been essential for progress in detector development for HEP over the past 
three decades and will be increasingly important as channel density and rates increase and environmental conditions become more challenging.

By semiconductor foundry standards HEP is a small-volume, high-mix customer. One of the mainstays for access to foundries by the HEP community 
has been the Metal Oxide Semiconductor Implementation Service (MOSIS), operated by USC. MOSIS, founded in 1981, provides access to 
state-of-the-art (SOTA) and state-of-the-practice (SOTP) foundry processes at nodes from 500 nm down to 12 nm using multi-project wafer runs so 
that multiple parties can share the cost of mask sets and wafer processing during prototyping. It should be noted that while MOSIS provides good 
access to SOTA CMOS processes, it does not provide access to other important processes such as III-V, superconducting, specialty 
back-end-of-line processing such as back thinning and thin entrance windows, or integration (2.5D and 3D). In many cases “prototype” runs, which 
can produce 500-2000 die, are sufficient to support small to mid-sized HEP experiment needs. For larger experiments, a dedicated “engineering run” 
typically suffices.

The HEP community needs to maintain access to both SOTA and SOTP CMOS foundry processes. Access not only includes wafer processing, but access to 
design tools and data products -- Intellectual Property (IP) blocks, Process Design Kits, device performance data etc. In addition to CMOS 
electronics, foundry access is needed for other silicon devices such as CCDs that require somewhat different processing. For CCDs Lawrence 
Berkeley National Laboratory (LBNL) and FNAL have successfully transferred the technology to a new commercial vendor, Microchip, to replace 
Teledyne DALSA when they elected to no longer produce these detectors. Working devices have also been produced with MIT-LL.

Very recently a disruptive access model with open-source, non-disclosure-agreement (NDA)-free ASIC prototyping has been spearheaded by the Google 
and the Skywater foundry. Multiple HEP groups are exploring this option for future development and further engagement and early adoption of open 
source design, where possible, should be encouraged. This is not only a potential way to reduce prototyping costs, but more importantly a way to
involve physics students (as ooposed to specialized electrical engineering students) in ASIC design.

Potential synergies in the context of the recently passed CHIPS and Science Act~\cite{CHIPSact} with the semiconductor detector program need 
to be explored in the near future.

Superconducting sensors are a key enabling technology for many HEP experiments with advances in sensor capabilities leading directly 
to expanded science reach. The unique materials and processes required for the fabrication of these sensors makes commercial sourcing 
impractical in comparison with semiconducting devices. Consequently, the development and fabrication of new sensors are often 
performed at academic clean rooms supported through HEP basic detector research and/or project funds. While this operational model 
has been successful to date, we are at a turning point in the history of superconducting electronics, as evidenced by the rapid 
growth in the field of quantum computing, when scale and sophistication of these sensors can lead to significant progress. In order 
to achieve this progress and meet the needs of the next generations of HEP experiments, it is necessary to broadly support all stages 
of the superconducting sensors development and to support a dedicated facility for this technology that is focused on HEP 
applications (with broader connections to non-HEP needs and industry).

\subsubsection{TDAQ Facility}
The creation of a dedicated (distributed) R\&D facility that can be used to emulate detectors and TDAQ systems, offers opportunities 
for integration testing (including low- and high-level triggering, data readout, data aggregation and reduction, networking, and storage), and 
developing and maintaining an accessible knowledge-base that crosses experiment-project boundaries.

\subsection{Synergies}
\subsubsection{Collaborative Research Models}
The US HEP model for these has been to base facilities and core capabilities at National Labs as user facilities funded by KA25. Alternative 
approaches might be considered based on models such as the CERN RD research programs or the NNSA NA-20 research consortia. The CERN RD programs 
provide an organizational framework within which collaborative research is conducted, but provide no direct funding support. However, engagement in 
these RD programs often provides a strong argument for grant funding to support participation. In contract, the NA-20 research consortia are 
multi-year multi-institutional research proposals with funding provided to the university groups and an expectation of support from National Labs 
already funded by the office. In both cases, the intent is to build 5+-year programs of research to address larger programmatic ambitions than 
individual institutions could tackle. The HEP investment in LAPPDs was to a large extent such a program.

Recently, both the DOE Instrumentation BRN~\cite{osti_1659761} and the ECFA Detector Roadmap (2021)~\cite{Detector:2784893} noted the inherent value 
of the CERN RD collaborations to instrumentation R\&D. The latter recommended the strengthening of existing RD collaborations and the creation of 
new ones in Calorimetry, Photo Sensors and PID, Liquid Detectors and Quantum Sensing. These RD collaborations are proposed to be global in extent and 
could be hosted by labs around the world. This plan has been approved by CERN Council and ECFA have been charged with its implementation over the 
next several years. We recommend the US HEP community engage broadly and early to help shape the global detector RD collaborations being established 
by ECFA and benefit from them.
 
Funding constraints in the US instrumentation R\&D program limit the ability to establish detector R\&D consortia. We recommend a significant 
increase in funding to enable DOE-HEP to establish a funding mechanism for collaborative or consortium funding. It should 
be understood that these research consortia will be multidisciplinary in nature and must include funding for scientists and engineers outside of 
physics.
 
\subsubsection{Multidisciplinary Research}
The topic of multidisciplinary R\&D was explored in the MultiHEP 2020 workshop~\cite{multihep2020}. This consisted of a series of invited presentations and panel 
discussions to collect experience on successes, challenges, and lessons about multidisciplinary collaborations to further HEP  
R\&D. The focus is on HEP developments that make use of expertise from outside HEP through collaborative efforts. R\&D disciplines explored 
included chemistry, materials science, nuclear science, micro and nano fabrication. Examples included small efforts at universities, projects 
at National Labs, work with industry through the DOE's Small Business Innovation Research (SBIR) program and otherwise, long-term multi-institute 
development such as LAPPD, and agency experience. The workshop panel discussions identified relevant points in the categories working with other 
fields, working with industry, and multidisciplinary personnel and funding.

\paragraph{Working with Other Fields}
Development of new instrumentation is often only possible through expertise and capabilities from outside HEP. Examples at the workshop were 
LAPPD, barium tagging for 0$\nu\beta\beta$ decay, use of carbon nanotubes for photon detection, etc. Informal discussions are required to 
establish the work scope for reasons summarized as ``because you don't know what you don't know''. Some interest in the HEP goals is more 
important than the expert's standing in the non-HEP field. The research philosophy is different in other fields. In HEP negative results are a 
success (most HEP results are negative), while other fields have narrowly defined goals and positive results are the measure of success. The 
publication approach is also different in other fields. There are no pre-prints since scoops are common and thus results remain secret until 
published. Publications are geared towards the next funding proposal. Author lists are small. It is a good idea to have agreements formalizing 
collaboration, funding, author lists etc. after the initial informal discussions.

\paragraph{Working with Industry}
There were numerous reports about successful SBIR efforts. The SBIR program offers a good path to fund multidisciplinary development by partnering 
with an appropriate firm. True partnerships between HEP scientists and firms are essential for success. An issue with SBIR is that many 
technologies critical for HEP do not have a lucrative commercialization path, but SBIR requires commercialization as the end product of Phase 2. 
Even when commercialization is in principle possible, there is a large gap between Phase 2 and volume production for profit. It is difficult for 
small companies to bridge that gap. 

Connecting with the right company is a challenge. Networking efforts can help. NP holds an annual exchange meeting for this purpose. Some National 
Labs have industry days, but not regularly or at the same level. BNL has a discovery park incubator. 

For multidisciplinary work there may be co-founding opportunities. Each office has to allocate and spend the required SBIR levels and this leads to 
round-off errors because each office has to spend down an exact amount. This would allows NP to co-fund something with HEP, for example.

\paragraph{Multidisciplinary Personnel and Funding}
Very few institutes train people in multiple disciplines at once. It is therefore hard to recruit people to carry out multidisciplinary 
work. On the other hand, a student or postdoc developing in two disciplines at once may have a hard time finding the next job, because 
they will not be competitive in either one of these disciplines with peers who devoted themselves fully to it. They need to find a job 
where both disciplines are required. Funding agency offices tend to not like paying for personnel normally under other offices, which 
makes it hard to find multidisciplinary work with a single grant. Recent QIS and micrelectronics Funding Opportunity Announcements (FOA) 
have aimed at multidisciplinary work. Microelectronics has been more successful because (unusually) multiple offices co-funded the same 
FOA, while this was not the case for QIS, where each office had separate FOAs. More funding like microelectronics is needed if fostering 
multidisciplinary work is desired. 

\subsubsection{Co-Design Approach}
Most past and current detectors were integrated as parts into the whole with assumptions about on- and off- detector communication and 
monitoring coming as either afterthoughts or as separate sub-system interface points. The complexity of the detectors for future 
experiments will require a more cooperative detector level co-design and co-development atmosphere. The community has the tools 
and repositories capable of creating and archiving an evolving detector design with sub-detector sections and interface documents 
created as functional blocks in separable areas with the hooks to create a coherent overview. An ideal system repository would hold 
representational blocks for fledgling sub-systems under development with interface documents that include target specifications such 
as data transmission protocols. 

The full detector level framework for high-level simulation should be developed for each sub-system to participate in a functional 
simulation with abstracted performance parameters tuned using guidance from lower-level test bench simulations that are compatible with 
its internal design. In this way operational modes such as calibration, monitoring and data-taking, can be tested across system 
boundaries. This will help expose easy-to-ignore but important interface details that must be coherent across systems and can evolve to 
provide realistic latency and throughput for each sub-system that will be a basis for validation of intra-sub-system communication fidelity. 

In the case of detector-specific simulation and reconstruction software, the tradition is that the detector and software design efforts 
are separate processes, with the latter dealt with as an afterthought. Consequently, suboptimal choices in terms of detector technology or configuration may lead to computationally expensive simulation and reconstruction algorithms, resulting in a longer publication turnaround 
time, physics results of lower quality, or the need to increase investment in computing resources. A collaboration model that includes the 
co-design of the detector and its software would allow to optimize physics performance simultaneously with the cost of hardware and 
computing~\cite{Krutelyov:2016wxl}.

\subsection{Workforce Development for Instrumentation}
The instrumentation workforce includes not only the physicists making the connections to the particle physics science mission, but the world-class 
scientists from other disciplines and the uniquely skilled technical staff that make the enterprise a success. Engaging with world-class scientists 
from other disciplines requires not only funding, but rewards and recognition commensurate with their contributions to the success of the HEP program. 
This extends to the technical staff without whom no major detector program could be realized. A summary of the topic can be found in~\cite{https://doi.org/10.48550/arxiv.2204.07285}.

There are two cultural hurdles that need to be addressed. The first is embracing scientists from other disciplines as equals. The second is the issue 
that both the scientists from other disciplines and the technical staff are funded primarily from soft funding streams (e.g. projects). This makes it 
very difficult to sustain staff expertise through lulls between major construction projects and makes these positions less attractive than comparable 
positions in industry that typically also come with greater financial compensation. Establishing a set of hard funding opportunities that allow for 
key technical positions to be funded as career opportunities could attract more highly qualified candidates for these vital roles. 

\subsection{Blue-sky Research and Tackling Grand Challenges in Instrumentation}

Innovative instrumentation research is one of the defining characteristics of the field of particle physics. Blue-sky research is defined as highly 
innovative research that does not have a predefined application and is typically characterized by high risk and potentially high reward. In contrast, 
Grand Challenges have a defined goal but are transformational rather than incremental in nature. The Detector BRN defines a number of Grand Challenges 
for detector instrumentation.

In particle physics these R\&D activities have often been of broader application and had immense societal benefit. Examples include: the development 
of the World Wide Web, Magnetic Resonance Imaging, Positron Emission Tomography and X-ray imaging for photon science. It is essential that adequate 
resources be provided to support more speculative blue-sky R\&D which can be riskier in terms of immediate benefits but can bring significant and 
potentially transformational returns if successful both to particle physics (discovering new physics may only be possible by developing novel 
technologies) and to society.

\section{Summary}
Advances in detector instrumentation enable advancements in HEP. The US has historically been among the leaders in the field in many areas of 
technologies. However, the shortage of the detector R\&D program in the US has become evident, such as the opportunity for sustained 
careers in instrumentation, from student to tenured scientist or faculty, or the highly impactful CERN RD collaborations. Investments have to be 
made to close the gap in these areas and ensure that US HEP detector instrumentation will continue to be at the forefront of technological 
innovation and scientific breakthrough. The Snowmass Instrumentation Frontier has identified strategic R\&D directions that would be beneficial to 
pursue within each of its Topical Groups. For a more detailed discussion we point the reader to the Topical Group reports. Furthermore, we have 
identified organizational and strategic needs that are common to a large part of detector technologies for HEP applications, such as common 
facilities, research consortia, funding models or multidisciplinary work. This process has also identified the need for regular funding opportunities 
for small-scale experiments, which typically drive detector innovation on a much shorter timescale than the few mega-projects that comprise the 
majority of the HEP budget. Such innovation in turn benefits the entire program at large. 

\bibliographystyle{JHEP}
\bibliography{Instrumentation}